\documentclass[preprint,article,accept,pdftex,moreauthors]{Definitions/mdpi} 

\usepackage[numbers]{natbib}
\firstpage{1} 
\pubvolume{1}
\issuenum{1}
\articlenumber{0}
\pubyear{2022}
\copyrightyear{2022}
\datereceived{} 
\dateaccepted{}
\datepublished{} 
\hreflink{https://doi.org/} 
\pdfoutput=1

\Title{Statistical Issues in Serial Killer Nurse Cases}

\TitleCitation{Statistical issues in Serial Killer Nurse cases}

\Author{Richard D. Gill$^{1\ast}$\orcidA, Norman Fenton$^{2}$\orcidB{} and David Lagnado $^{3}$\orcidC}

\AuthorNames{Richard Gill, Norman Fenton and David Lagnado}

\AuthorCitation{Gill, R.D.; Fenton N.; Lagnado, N.}

\address{%
$^{1}$ \quad Mathematical Institute, Leiden University, Leiden, The Netherlands; gill@math.leidenuniv.nl\\
$^{2}$ \quad Queen Mary University of London, London, GB; n.fenton@qmul.ac.uk\\
$^{3}$ \quad University College London, London, GB; d.lagnado@ucl.ac.uk}

\corres{Correspondence: gill@math.leidenuniv.nl}

\abstract{We study statistical aspects of the case of the British nurse Ben Geen, convicted of 2 counts of murder and 15 of grievous bodily harm following events at Horton General Hospital (in the town of Banbury, Oxfordshire, UK) during December 2013--February 2014. We draw attention to parallels with the cases of nurses Lucia de Berk (the Netherlands) and Daniela Poggiali (Italy), in both of which an initial conviction for multiple murders of patients was overturned after reopening of the case. We pay most attention to the investigative processes by which data, and not just statistical data, is generated; namely, the identification of past cases in which the nurse under suspicion might have been involved. We argue that the investigation and prosecution of such cases is vulnerable to many cognitive biases and errors of reasoning about uncertainty, compilcated by the fact that fact-finders have to determine not only whether a particular person was guilty of certain crimes, but whether any crimes were committed by anybody at all. The paper includes some new statistical findings on the Ben Geen case and suggests further avenues for investigation. The experiences recounted here have contributed to the writing of the hand-book Green et al.~(2022), \emph{Healthcare Serial Killer or Coincidence? Statistical Issues in Investigation of Suspected Medical Misconduct}, commissioned by the Royal Statistical Society, Statistics and the Law section. Submitted to \emph{MDPI Laws}. \textbf{This version: 5 August, 2022}}


\begin{document}

\tableofcontents

\section{Introduction}\label{Introduction}

Serial murder by health care professionals is, fortunately, very rare.
The investigation and prosecution of suspected cases are very complex.
Fact-finders must not just determine who was the perpetrator of a
string of crimes -- they must also determine whether any crimes
occurred at all. Clusters of unusual events in a health care context can
arise for numerous reasons, which can initially be completely unsuspected.
They can become associated with the presence of a particular nurse 
for wholly innocent reasons, as a number of well documented cases proves.
Medical evidence of wrong-doing might be ambiguous or on its own,
inconclusive. Statistical evidence is often brought to bear, sometimes through
explicit probability calculations; but in other cases the raw numbers appear to speak
for themselves, and professional statisticians are not involved (or worse still, they
are ignored). High weight is given to psychological evaluations and to  
character judgements of colleagues at the workplace, and to striking
but possibly ambiguous ``smoking gun'' events.
The legal community, as well as the public, have little understanding
of statistics and probability. Such cases are of course extremely shocking and
attract great media attention. The public will often draw their own conclusions
on the basis of a sinister interpretation of one or two odd events.

The lead author of this paper (Richard Gill) has now been professionally involved in three such cases:
the cases of the Dutch nurse Lucia de Berk, the British nurse Ben Geen, and the Italian nurse
Daniela Poggiali. The case of Lucia de Berk is very well known and a number of scholarly 
publications have been devoted to it. The highly complex case is well summarized on English language Wikipedia,
\url{https://en.wikipedia.org/wiki/Lucia_de_Berk}, and we will give further information about the case,
including literature references, later in the paper.
That case started with Lucia's arrest in 2001 and finished
with her acquittal at a retrial in 2010. Ben Geen was arrested in 2004, and within a few years he was 
sentenced with life imprisonment. Attempts have been
made by his lawyers to have the case reopened by the CCRC, but so far without success. 
The consensus on English language Wikipedia is that he is guilty, \url{https://en.wikipedia.org/wiki/Benjamin_Geen}.
The reader who was previously unaware of the case is recommended to start by looking at what is
largely the prosecution's view presented on Wikipedia.
Daniela Poggiali likewise was arrested in 2014. She spent a number of years in prison but was
acquitted in 2021 at the Italian court of cassation. A first scholarly publication on the case is \citep{dotto-etal}. Knowing the
Italian justice system, this might not be the last word on the case, as far as legal history is concerned.

The purpose of this paper is to draw attention to the statistical issues in the Ben Geen case,
establishing parallels between it and the case of Lucia de Berk, and to point out avenues for
further investigation. We believe the case deserves serious scholarly attention, 
and therefore ``kick off'' with analysis of statistical aspects (``statistical'' in a broad sense!) of the
case, which is of course where the authors' expertises lie.
The medical side of the case is crucial and deserves further attention too.
We plan to survey that part of the case in a subsequent publication. 
Lead author Richard Gill is at present personally convinced that Ben Geen is innocent,
not only because of the evidence presented in this paper but also thanks to familiarity with the
medical and psychological evidence, as well as personal contacts with various main players in Ben Geen's trials.
In Gill's opinion, the medical evidence is highly ambiguous, and the jury (like the media) was swayed by
the psychological and circumstantial evidence presented in the case, which provided the prosecution with
a psychologically compelling story (both simple and shocking) which however did not do justice to the facts.

Gill is also a co-author of an  upcoming publication of the Royal Statistical Society (RSS)
entitled \emph{Healthcare Serial Killer or Coincidence?
Statistical Issues in Investigation of Suspected Medical Misconduct}, \citet{green-etal}.
That work is a booklet or small handbook explaining statistical issues and giving advice to all parties involved in criminal investigation and prosecution
of this kind of case: hospital authorities, medical specialists, nursing experts, police investigators, prosecutors, defence lawyers,
statistical and medical expert witnesses,
journalists, and last but not least, the public.
It was commissioned by the \emph{Statistics and the Law} section of the RSS and has been peer reviewed in that section\footnote{Authors of the booklet are 
Professor Peter Green FRS, Emeritus Professor of Statistics, University of Bristol, and Distinguished Professor, University of Technology, Sydney (chairman); 
Professor Richard Gill, Emeritus Professor of Statistics, Leiden University;
Neil Mackenzie QC, Arnot Manderson Advocates, Edinburgh;
Professor Julia Mortera, Professor of Statistics, Università Roma Tre;
Professor William Thompson, Professor Emeritus of Criminology, Law, and Society; Psychology and Social Behavior; and Law, University of California, Irvine.
One of the appendices was provided by Professor Jane Hutton, Professor of Medical Statistics, University of Warwick.}.
Under ``statistical issues'' we do not only mean technical matters concerning the correctness or interpretation of formal
statistical methods, but also the dangers of biased data gathering, and the danger of cognitive biases in reasoning about evidence.
These issues are equally relevant in what are called common law juridisctions where investigation and adjudication 
is based on adversarial principles, and in what are called civil law jurisdictions where an inquisatorial system is employed.
The ``fact finder'' is often a jury of ordinary citizens in the former case, and a judge or board of judges in the latter case.
However, in the case of adversarial systems, a judge often plays a major role in determining
what evidence may be brought to the attention of the jury, and may advise how that evidence should be interpreted by 
the jury. This RSS booklet provides the necessary background material on the general legal and statistical issues covered in more detail in this paper.

The paper is organised as follows. After the present section 1 ``Introduction'' comes the obligatory Section~\ref{materials}) on \emph{Materials and Methods},
very briefly describing data sources and data sets and statistical (and other) software which will be used when looking at the two specific
cases of Lucia de Berk and of Ben Geen. In Section~\ref{lucia}, we briefly look at the statistical issues in the
case of Lucia de Berk. The next two sections consist of a general overview of the Ben Geen case (Section~\ref{geen}), and
then a presentation of some new statistical findings (Section~\ref{geen-plus}), which we think deserve further investigation and which
might be useful in future attempts to reopen the case.

Section~\ref{criminology} discusses the literature of criminology and forensic psychology on health care serial killers. 
The key issues of evidence interpretation discussed in this section are then modelled using Bayesian networks in Section~\ref{bayesian}. 
In Section~\ref{confirmation} we discuss confirmation bias from the point of view of cognitive psychology, as well as referencing the
medical statistical and epidemiological literature on statistical biases in observational studies, and tactics for avoiding or mitigating them.
In Section~\ref{conclusions}, we draw some final conclusions and make recommendations for further research.

\section{Materials and Methods}\label{materials}

The materials of this study are data sets provided to the authors
by the defence lawyers of Lucia de Berk and Ben Geen. In the former case
they were originally submitted to police investigators by the hospitals where de Berk had worked.
In the latter case, they were obtained through FOI requests 
made by Geen's lawyers to numerous hospitals. All data files may be obtained from the corresponding author Richard Gill.
Statistical analyses are carried out using the open source general statistical system ``R'', and the 
Bayesian network software for risk analysis and decision making ``AgenaRisk''.
The data files and programming scripts which generate the statistical graphics presented 
in the paper can be found in the ``github'' repository \url{https://github.com/gill1109/bengeen}
or directly from the corresponding author. He can also provide extensive files on the medical
evidence on the 18 patients whom Geen was accused of harming.

\section{The case of Lucia de Berk}\label{lucia}

The case of the Dutch nurse formerly known as ``Lucia de B.'' starts
with three \(2 \times 2\) tables, see Figure
\protect\hyperlink{figure1}{1}.

\noindent\protect\hypertarget{figure1}{}{}\includegraphics[width=14cm]{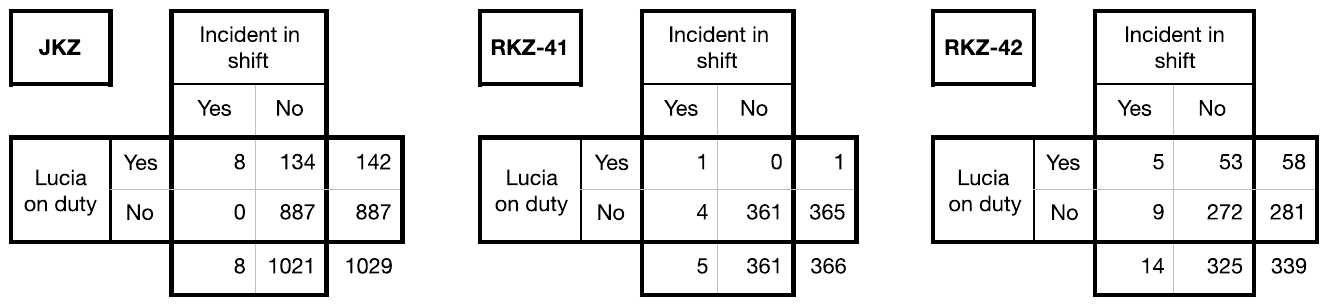}

\noindent\textbf{Figure 1}. Roster data from the case of Lucia de B.

\bigskip

Between November 2020 and September 2021 (when police investigations
started) on a medium care ward at the Juliana Children's Hospital
(``JKZ'') in the Hague, there were in total 1029 (\(3 \times 343\))
8-hour shifts (3 shifts a day, 7 days a week). In 8 of them, an
``incident'' occurred. All 8, in the shifts of a nurse
Lucia de B. Several years earlier, at another hospital, the Red Cross
Hospital (``RKZ"), during the same four months in two intensive care
wards (Wards 41 and 42; the data from ward 42 misses 9 days at the
beginning and end of the 4 month period) there were 5 and 14 incidents
respectively. Lucia was only on duty once in RKZ-41 but on just that one
occasion she netted one of the 5 incidents. She mainly worked on Ward
42, where she netted disproportionately many of the 14 incidents (she
had one third of the incidents in only one sixth of the shifts).

The data can also be found in a recent paper \citep{gill-etal}. 
These particular numbers were statistically analysed and interpreted for the court in
2002 by law psychologist Henk Elffers, who had been contacted in 2001 by police
investigators at the beginning of what became a ten year saga. He later 
kindly made his reports to the court, \citep{elffersa, elffersb}, 
available for scientific research (English translations prepared by Gill).
In the reports, Elffers deduced that the chance that an innocent nurse could have
experienced as many incidents in her shifts in all three wards as Lucia did,
under the assumption that incidents and shifts were independent of one another, 
and the data from the three wards too, was 1 in 342 million.

Elffers' probability of 1 in 342 million was widely reported in the media. 
It had an enormous impact  on everyone who heard it, convincing seasoned and sceptical
journalists, who till then had not found the evidence very convincing at all,
of Lucia's guilt. Elffers had actually made an elementary methodological error, multiplying three p-values
and interpreting the result as another p-value, an error which only became evident
several years later, after his reports had been made public. The defence had
recruited two philosophically inclined mathematicians to give counter evidence,
but, apparently equally ignorant of standard statistical methodology, 
they were unable to calculate any p-value at all, and argued
that it was principally impossible to come up with any particular probability. 
Elffers could, and he could convince the judges that they understood his analysis.
Had the experts for the defence applied a simple textbook analysis (Fisher's method of combining independent p-values)
they would have come up with a p-value of 1 in a thousand,
even while accepting the data and accepting Elffers' mathematical assumptions.

Accepting the data for the moment, does one need a statistician to interpret it to the board of judges of a
criminal court where Lucia is being tried for serial murder? The data do also
speak for themselves. This was the opinion of the judges at
Lucia's appeal in 2004: they wrote at the beginning of the motivation of their final verdict
``a statistical calculation of probability plays no part at all in our [guilty] verdict''. 
Lucia was convicted by them ``solely on the grounds of
irrefutable and scientific medical evidence''. 
In fact, their more than 100 page published justification shows that they, 
and many medical experts too, were already convinced \emph{because of the raw
statistics} that Lucia's presence at so many incidents could not be
chance; her mere presence was explicitly the reason that several medical experts classified
certain incidents as \emph{inexplicable} and hence \emph{suspicious}.
Toxicological evidence concerning one death was then sufficient, by a so-called chain argument
(which was in fact an informal Bayesian argument), to turn other deaths
and resuscitations into ``scientifically irrefutably proven'' murders
and murder attempts.

The judges considered Lucia to have been a particularly refined murderer since she
continued to to assert her innocence, and had been so cunning as to leave 
no evidence at all of \emph{how} she had killed many of the patients of whose
murder she was found guilty by the court.

Do the data speak the truth, the whole truth, and nothing but the truth?
No. Lucia's life sentence for ten murders of children and old people got
reversed. At the retrial in 2010, the judges, in their spoken summing up,
congratulated all the nurses including Lucia on their devotion and their professional
efforts to save the lives of their patients, lives which (as the judges now
announced) had been unnecessarily shortened through medical errors.
The errors were caused by mis-diagnosis, chaotic management, ignorance of
the content of transferred patients' medical dossiers. They were
committed by hospital specialists and hospital managers, as is
abundantly clear from the medical evidence given at the re-trial.

The key to Lucia's exoneration was the invalidation of the toxicological
argument concerning the ``trigger-case''. We now know that a coincidence
of several triggers set off a witch-hunt, followed (as was already noted
by observers at the time) by what seemed like a witch-trial. A nurse who
stood out from the crowd with a striking appearance, a strong
personality, and a colourful (dark?) past was a natural scapegoat for
the mistakes being made in a failing hospital department.

We now also know that the original Lucia case numbers (Figure 1) were
wrong. Events have been misclassified. Some have
been opportunistically shifted from one shift to an adjacent shift. Quite a few have been
omitted -- they weren't ``unexplained'', so they weren't included as
``incidents''. Several real incidents (i.e., ``incidents'' according to
the law) have been suppressed. They should have been reported to the
health inspectorate, but this wasn't done. For instance, one incident
was actually a case of euthanasia deliberately performed by the medical
doctors of a child with severe birth defects, which, illegally, was not
reported as such to the inspectorate. One incident initially attributed to Lucia was later
found to have taken place during her vacation: it was re-classified as a non-incident.

As Elffers wrote in his second report: these numbers didn't arise by chance. He stated
that of course this did not prove that she was a murderer. He explicitly offered the
court three more alternatives: Lucia was maybe just a bad nurse; maybe she had harder shifts than her colleagues;
maybe she was set up by someone else. In their summing up, the board of judges explicitly showed why
they discounted the first two of the three alternatives. They paid no attention to the third,
and the defence had not raised that possibility either.

The data from JKZ was actually compiled in a great hurry by persons
who were already convinced they were dealing with a
serial killer and embarrassed that they didn't catch her earlier. Lucia
was in fact set up, in the sense that the hospital's clinical director
(the head paediatrician) was already suspicious of her and was 
waiting for a final event to clinch the case she had been building up; 
she had already compiled the dossiers of
selected earlier events.

Why, and how, did Lucia get a re-trial? Answer: she
was very lucky. There was an almost-inside whistleblower. The
sister-in-law Metta de Noo of JKZ's chief paediatrician Arda Derksen-Lubsen 
was a medical doctor, knew her sister-in-law very well, and moreover, 
Metta's son and daughter, studying medicine in Leiden, lived with their uncle and aunt
in the Hague. Even so, it took a couple of years for the penny to drop, that there was
a lot wrong with the medical evidence being used in the case.
Metta de Noo recruited her brother, philosopher of science Prof.~Ton Derksen.
Together the brother-sister pair wrote a book devastatingly analysing the court's arguments, 
gained media attention and a strong popular following especially in the academic world.
They prepared and submitted an application to the CEAS, the then Netherlands equivalent of the British CCRC.

There were more lucky breaks. Lucia had first experienced some
definite bad luck (though nothing like the 1 in 342 million which hit
the newspapers) but later, new coincidences put some of the best
medical and legal minds in the Netherlands in key positions along the tortuous route to a
re-trial. 

On the way, it was important to explain both to the public and to legal authorities what
had been wrong with the original statistical analysis, since its results had been so widely
disseminated and had had so much impact. Gill gave evidence to a
judicial review committee investigating the possibility to ask the supreme court of the
Netherlands to have the case reopened. The fact of the matter is, popular
unrest about an unsafe conviction seems to be necessary before authorities
will act on it. In the social media sphere, the scientific community was
activated and could well understand that the legal systems had not
been able to properly interpret hard scientific evidence whether from
statistics or from toxicology. The CEAS had indeed been set up because of some
recent miscarriages of justice caused in part by misinterpretation of scientific evidence.
The international reputation of the Netherlands' legal system
was at stake, as well as the Dutch people's trust in their system. Whether or not key persons believed that Lucia was innocent or
guilty was irrelevant: they did understand that justice had to be seen to be done,
and that the case was complex and controversial and needed to be handled with
the utmost care.

\section{The case of Ben Geen}\label{geen}

The data presented in Figure \protect\hyperlink{figure2}{2} played a role in getting
the young English nurse Ben Geen a life sentence for two counts of
murder and 15 of grievous bodily harm (a 16th count of grievous bodily
harm was not considered proven), in the three consecutive months of
December 2003, January 2004 and February 2004. We will describe the
case from the point of view of the defence, and focus on statistical
issues, though of course an enormous amount of medical evidence was deployed
by the prosecution too. However, perhaps the most crucial piece of evidence
(one could call it: the smoking gun) was a striking incident involving a syringe,
for which defence and prosecution had completely different interpretations.

\noindent\protect\hypertarget{figure2}{}{}\includegraphics[width=14cm]{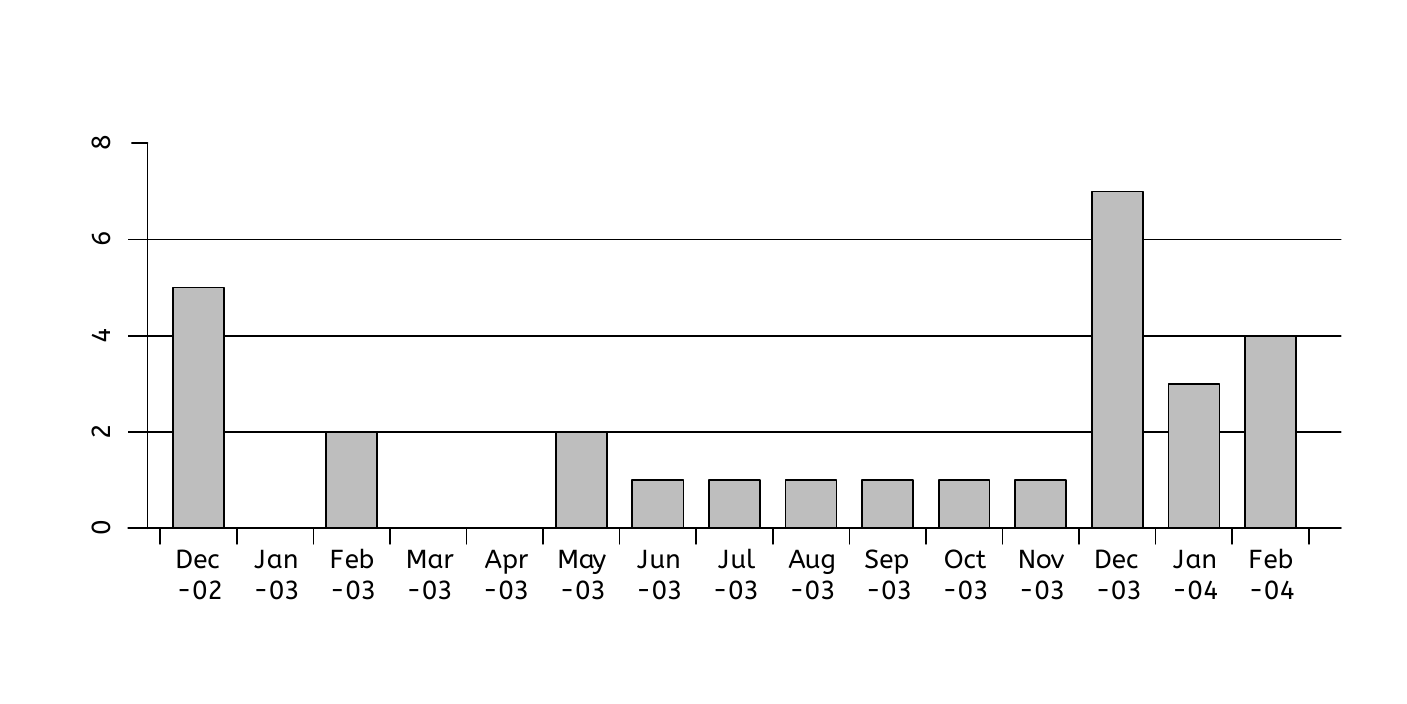}

\vspace*{-0.7cm}

\noindent\textbf{Figure 2}. Admissions to critical care from the emergency department with
a diagnosis of cardio-respiratory or respiratory arrest or
hypoglycaemia, data: Head Nurse Brock

\bigskip

There are many similarities between the investigation that led to Geen's
conviction and the erroneous investigation that led to Lucia's wrongful
conviction. Ben Geen was a striking figure, and the hospital he worked
at was in serious difficulties, leading to overworked staff and chaotic
administration. Some odd events made hospital authorities suspicious
of him. A very rapid and intensive search was made for evidence which might implicate him
in various events in the emergency ward where he worked, resulting in a
huge dossier which was passed to police. A background factor in
the case could well have been the so-called ``Shipman effect''. 
The Ben Geen case occurred shortly after the Shipman Enquiry, which
blamed health-care administrators for not earlier noticing serial killer
doctor Harold Shipman. Some speak of ``Shipman hysteria''.

Learning from mistakes is good, but a new danger then arises that by
learning the wrong lessons from one kind of mistake, one might increase
the chance of making the opposite mistake. If the rate of false
convictions goes down but nothing else changes, the rate of false
acquittals will go up. The more easily a health-care system goes into
alarm-mode because of suspicion that it harbours a health-care serial
killer, the more often innocent health-care professionals will trigger
an alarm.

The key data-set in Figure 2 was presented as evidence to the court by
Michelle Brock, head-nurse of the Accidents and Emergency department
where Ben Geen worked, at Horton General Hospital. This is a small
hospital in the provincial market town of Banbury in North Oxfordshire.
Together with a dossier of perhaps 30 incidents all from December 2003
onwards, it had initially been compiled in great haste before the case
was reported to the police. Michelle and some colleagues based their
work on patient records and nurse attendance records at the hospital,
\emph{looking only at what happened during Ben's shifts}. Their
investigation appears to have been triggered and guided by recent memory
and gossip. Ben, who was a trainee nurse, had won a higher qualification
at the beginning of December, allowing him to work under less
supervision than before. The final trigger for their investigation had
been two sudden and surprising collapses of patients who had just
entered Accident \& Emergency (A\&E, also often referred to as ED:
Emergency Department) on Thursday 5 February. Ben had reported sick on
Friday, and had had a free weekend after that. He was arrested on Monday
evening, 9 February 2004, when he came in to work to do a night shift.
That is only one third of the way into the last bar of the bar-chart.
The bar-chart was also known to the medical experts who were consulted
on the 18 individual cases. There seems no doubt that it had
impact on everyone involved in the trial, including journalists covering
the trial.

We catch a glimpse from the chart of the fact that a lot of old people
and people with existing serious health problems get brought to
emergency care during the winter months of December, January and
February with acute problems involving heart and lungs (a hot summer is
also a danger period). A common diagnosis is cardio-respiratory arrest
(the heart has stopped working and consequently the lungs too), much
less common is ``pure'' respiratory arrest (the lungs have stopped
working); fairly common is hypoglycaemia: a fall in blood glucose level.
It causes fainting; breathing stops or is much suppressed. It can be
caused by too much insulin or other glucose lowering diabetes tablets,
delaying or missing a meal, not eating enough carbohydrate, unplanned
physical activity, more strenuous exercise than usual, drinking alcohol
-- the risk of hypoglycaemia increases, the more alcohol you drink. In
the bar-chart, nurse Brock has combined the three ``standard''
categories cardio-respiratory, respiratory, and hypoglycaemic arrest;
but what is the correct category is hard to guess when a patient
presents (arrives at the hospital). Past medical history, and future
medical events will give clues as to what was actually going on. In an
emergency situation, past medical history may be unknown.

Hospital nurses and authorities had been worried by the behaviour of the
young male trainee nurse Ben Geen already before December 2003. His
father was in the army, his mother was a nurse. He had been in the
territorial army medical corps, and his ambition was to be qualified and
then transferred as a combat medic to a military field hospital in Iraq.
He was energetic and very ``present'', keen to get action and to get
experience. He made some other nurses nervous. They were calling him
``Ben Allitt'' behind his back, not a kind nick-name, since Bev Allitt
is the very well-known name of a pre-Shipman famous English convicted
serial killer nurse.

In December 2003 the numbers of patients reaching an overstressed
emergency ward in a small underfunded provincial hospital, threatened by
closure, probably understaffed, probably lacking resident consultants in
critical specialisms, was bigger than ever. There were two
``surprising'' events when patients who were initially thought to be in
fairly good shape suddenly, and at the time unexplainably, worsened. Ben
was usually around when anything happened simply because he was usually
around: he was working double shifts in order to gain more and more
experience as fast as possible, and also often filled in for absent
colleagues.

On Thursday 5 February 2004, at the end of an exceptionally hectic day,
a chronic alcoholic diabetic presented himself in the hospital (referred
to ED by his doctor), on account of severe gastric pain and vomiting,
and suffering fainting fits. Ben took a blood sample. The patient
suddenly worsened and later had little idea what else Ben did to him.
Ben certainly inserted a \emph{canula} (a tube that can be inserted into
the body, for the delivery or removal of fluid or for the gathering of
samples) and the patient was transferred to the Critical Care
department, CC. The real problem for Ben came later: Ben went home with,
unknown to himself (he said), a used plastic needle-less syringe
containing some muscle relaxant in his nurse's smock. Such a syringe is
used to administer necessary medications, including a muscle relaxant,
through the \emph{canula} prior to inserting breathing and feeding tubes
into patients on the way to Critical Care. Ben stayed home sick on
Friday, and then had the weekend free. His girlfriend, another nurse,
doing the washing, told him off for this (she said) and told him to take
it back as soon as possible. So, on Monday evening -- with the syringe
in his coat pocket -- he was met by policemen as he entered the
hospital. In some panic (he said) he stupidly further emptied the
remaining contents of the syringe into his pocket. The Prosecution
claimed that Ben tried to harm patients by injecting them with this
stuff so that he could then play the hero, helping to resuscitate them -
the so-called ``Munchhausen by proxy'' syndrome.

Ben and his girl-friend gave conflicting reports of the colour of the fluid in the
syringe. In fact, the medication in question changes colour when exposed to air, 
from transparent to bright yellow. The prosecution argued that the discrepancy
showed that Ben also often used the same syringe to harm patients, and
that the worn markings on the syringe confirmed that it was his common weapon.
The defense discovered that the markings rapidly wear off if a syringe is
kept for a short while in a damp pocket.

At his trial, the Crown secured the services of a famous and experienced
expert (a highly distinguished professor of Anaesthesiology), who found
a number of the events highly suspicious; another confidently swore that
never in his long experience had he met with an \emph{unexplained}
respiratory arrest.

Experts would agree with this claim because they strongly believe that
all respiratory arrests are ``explainable''. The problem is that different experts often
give different explanations. Whether a collapse is diagnosed as
cardio-respiratory, respiratory or hypoglycaemic can be pretty
arbitrary. Perhaps only weeks later (if the patient recovers and more has been
learnt of their state of health and underlying conditions) might it
be possible to give a more confident diagnosis of the cause of events
in the past.

When either heart or lungs get into difficulty, the other
organ rapidly gets into difficulties too. Hypoglycaemic arrest
(critically low blood glucose levels) always involves breathing problems
(you faint when not enough oxygen is reaching your brain) and can
trigger further deterioration of heart and lung function. Reduced oxygen
levels affect brain, heart, lungs. Muscles burn oxygen, the brain burns
oxygen. All arrests are explainable, but the categories which are ticked
on forms in the patient's dossier and in the hospital's administrative
records may differ and may be revised in the light of later events. The
categories which tend to be chosen by nurses, doctors and administrators
may depend on who is doing it, and may show trends and jumps as time
goes by. Just one occurrence of an unusual diagnosis alerts people to
its existence, and they start seeing it every day: the well known
\emph{Baader-Meinhof phenomenon} or \emph{Frequency Illusion}.

At the time each had actually occurred, each of the 18 cases in the
criminal charges against Ben had been considered ``normal''. The last
two had surprised some people (certainly not all), but because of
earlier suspicion and gossip, they triggered an emergency weekend-long
internal hospital investigation, in which more than 30 dossiers of
patients who had in recent months gone through Emergency \emph{while Ben
was on duty} were combed through, resulting in a dossier of 18 cases to
hand over to the police on Monday. In fact the team had access to 4000
patient medical records but were not interested in what happened when
Ben was not there. Expert witnesses for the defence later described how
explainable each of the 18 was, although they were honest enough to
admit that some cases were too complex to come to any clear conclusion.
The prosecution had more expensive and more court-experienced experts
than the defence. The prosecution experts were of course specifically
hired to point out anomalies in each of the selected 18 cases, and
tended to be rather confident of their diagnoses. Prosecution experts
are ``instructed'' by the prosecution, defence experts are
``instructed'' by the defence. Experts will report what documents they
were given to study, and on which their ``opinion'' is based, but prosecution experts
also tend to receive a lot of further verbal information (much of it
hypothetical) about the case from police or prosecutors. That certainly
happened in this case.

Ben must have used a myriad of different techniques to cause all these
unexplained medical emergencies and in many cases the expert witnesses
called by the Crown in fact had conflicting ideas of what he might have
done; but they agreed that he `must have done something'. All of the 18
patients were very sick, and what happened to each was what you may well
have expected to happen in view of their existing severe and often
complex conditions. But sometimes developments are fast, you do not
``see them coming'', and so a sudden worsening takes some nurses or some
doctors by surprise. People, including Ben himself, did notice Ben often
being there when such events took place. He had said, and said it in
court again, that he thought he had been jinxed.

Ben's unemotional and careful account of what he could recall that he
had seen and done in each case (he had received military interrogation
training), the impression he gave that he knew the law better than the
lawyers, an eminent professor's categorical statement that he had never
seen an unexplained respiratory collapse in all his career, and the
smoking gun which was the syringe, together clinched the matter for the
jury. The judge, in his summing up made it very clear what verdict he
expected from the jury, who had sat through the presentation of
interminable complex medical evidence (in 18 cases).

Interestingly, the judge had earlier instructed the jury to consider each of
the 18 cases entirely on their own merits. This would seem to be quite a
meaningless injunction. One can only guess at the thinking of the jury, but
if the members of the jury were convinced by the incident with the syringe
that Ben was deliberately harming patients and lying about his activities,
then they would clearly tend to think that each case about which suspicions could be raised
was likely indeed another case where Ben was intent on harming a patient;
and the number of such cases could only have strengthened their conclusion. 
The prosecution argued that there had been an unlikely large number of a particular
and suspicious kind of event during Ben's shifts. Ben himself admitted that
the number was strange. Whether it really was excessive is still not known, since
we do not have data on his working hours, we do not have times of all the events,
we know next to nothing about events which occurred when he was not on duty.
We do have plenty of evidence that the classification of events was highly biased,
and this bias was initiated in the hospital, as we will see.

The defence employed two expert witnesses who each delivered a comprehensive report on the
problem of deducing the reality of an observed pattern from medical events in an observational
setting (medical statistician Jane Hutton, clinical physiologist David Denison; their reports can be obtained from Gill).
They were both definitive that a ``pattern'' could not be established without statistical analysis
and that it would require further information, and in particular, requires blinding of medical experts, 
who should evaluate whether or not events were suspicious without knowing whether or not Geen was present.
The judge did not allow this evidence to be presented in person by Hutton or Denison in the court,
and told the jury that it consisted merely of common sense. The pattern was evident: 18 cases
when Geen was present, and each should be judged on its own merits. No statistical evidence was ever considered
beyond the raw numbers presented by nurse Brock and 
the opinion of a notable anaesthesiologist that  ``unexplained respiratory arrest is exceedingly rare''.
In particular, no information was presented on Geen's working hours, and no comparison was
made of the rate of certain events inside and outside of his shifts.

Blood and urine samples from the trigger case showed traces of a muscle
relaxant as well as of plenty of sedatives, but unfortunately some of
the samples were not dated (for there was not even information on when
they were taken nor by whom). Sedatives and muscle relaxant should have
been present. The traces of muscle relaxant were of the same kind as was
in the syringe. The consultant anaesthesiologist who had attended to the
trigger patient in intensive care said that she had asked (another
nurse, later) for a different one. Ben said that he was not told to
administer muscle relaxant, so, of course, had not done so. Hospital
records were woefully incomplete. Since the earlier cases were not at
the time thought to be suspicious, samples of blood and urine had not
been taken or had long ago been thrown away.

The annual pattern we see in that data is typical of many similar
hospitals all over Britain, although (for reasons explained above) there
are no data whatsoever about unexplained respiratory arrests. The data
stored in a hospital database are administrative data. Every event has
been put into a pre-existing category with an explanation, because it is
not possible to enter it into the data base otherwise. The data in the
database determines the fees of the medical consultants (the medical
specialists) and the funding of the hospital. The data are not collected
for scientific research or forensic investigation.

The three standard categories relevant to this case are
\emph{cardio-respiratory arrest}, \emph{respiratory arrest}, and
\emph{hypoglycaemic arrest}. We already presented the data supplied to
the court by Ben's head nurse, combining those three categories. Much
later, Horton General hospital provided us with the data as presently
archived in official hospital records, see Figure
\protect\hyperlink{figure3}{3}.

\noindent\protect\hypertarget{figure3}{}{}\includegraphics[width=14cm]{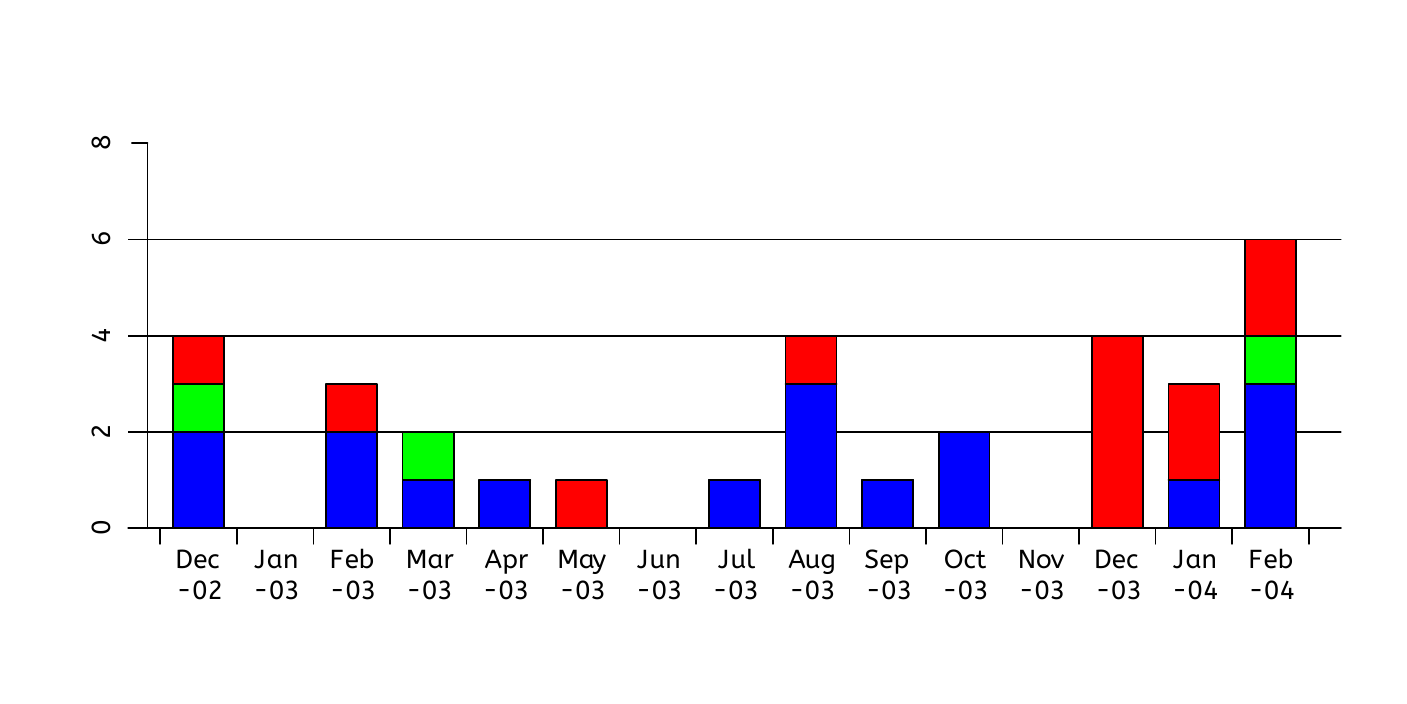}

\vspace*{-0.7cm}

\noindent\textbf{Figure 3.} Admissions to CC from ED with CR, Hypo or Resp arrest, FOI
data, Cardio-respiratory (blue), hypoglycaemic (green), respiratory
(red)

\bigskip

Critically, the data were now different and also (while the categories
are still separate), we have data from many more years. The total
numbers of relevant cases in December 2002 and in December 2003
(previously 5 and an incredible 7 respectively) are now equal -- both an
unremarkable 4. The split between categories in the two periods of
winter months is markedly different. In winter 2002 -- 2003 it is
normal, spread out over all three, but mostly cardio-respiratory. In
winter 2003 -- 2004 8 out of the 13 cases are categorised as
\emph{respiratory}. The total number of cases in January, in both
winters, is less than in adjacent months, which is normal.

Normal case-mix (for the three categories of interest), both in this
hospital and in all others (we have similar data from about 40 other
hospitals all over England, for the thirteen year period 2000 -- 2012),
is a mix mainly of cardio-respiratory, with respiratory and
hypoglycaemic normally each at roughly a fifth of the level of
cardio-respiratory. They are both much less usual, but neither can be
called \emph{rare}.

There are also data in the official public enquiry after Ben's
conviction, held to find out why he was not caught earlier and to
prevent such a tragedy from ever occurring again. The enquiry stated:
``The~number in December 2003 was six and this was~only one more than in
December 2002''. Two different numbers, yet again. The enquiry suggested
that the large numbers of incidents while Ben was carrying out his
attacks might have been expected anyway, due to the winter season,
thereby masking incidents caused by Ben. It did heavily criticise the
Emergency department for poor record keeping when updating patient
medical notes and very poor registration of patient drug
administrations.

\section{New findings in the Ben Geen case}\label{geen-plus}

We will now present some summary statistics based on data obtained from
Ben Geen's hospital which has been available for several years but never
looked at before. Instructed by Geen's defence team, Gill was asked
to answer certain questions about the ``normal situation'' in hospitals
\emph{like} HGH, using data that had been obtained from numerous FOI
requests to hospitals all over the country. He did his best to answer
exactly those questions and tried to maintain his scientific objectivity
by not learning about other aspects of the case, and therefore only later
took a close look at the data from HGH.

This means that Gill actually fell into a similar trap
as Elffers had in the case of Lucia de B.:
he tried to answer the question that had been put to him, without learning
enough about the context in order to judge whether it was a meaningful question.
In the Lucia case, Elffers just trusted that the hospital data had been compiled in a responsible manner.
In the case of Daniela Poggiali \citep{dotto-etal}, a breakthrough was made when statisticians took a look
at statistical aspects of the evaluation of toxicological evidence, following up a hint from one
of the medical experts for the defence. These cases are multidisciplinary; good multidisciplinary scientific
research requires communication between all parties. The legal context has a tendency for
communication of information to be highly restricted.

\noindent\protect\hypertarget{figure9}{}{}\includegraphics[width=14cm]{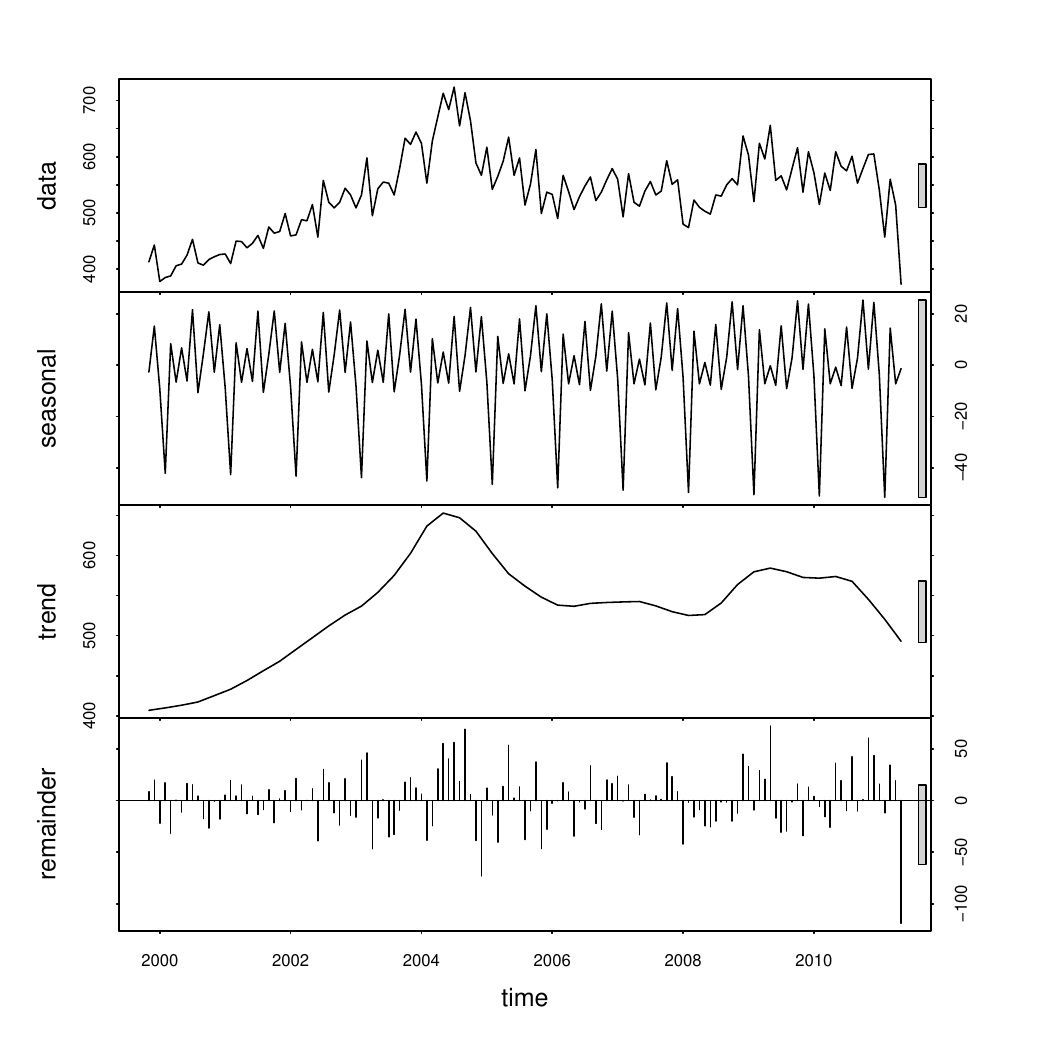}

\vspace*{-0.5cm}

\noindent\textbf{Figure 9.} Monthly Admissions to Emergency; decomposition of data into
trend, seasonal, remainder.

\bigskip

Let's take a look at the monthly total number of admissions in the
emergency room of his hospital over a thirteen year period roughly
centred on the critical end of 2003 -- beginning of 2004; see Figure
\protect\hyperlink{figure9}{9}. We have given this time series to the
go-to algorithm in the R package which, in an iterative procedure using
a moving window of length 21 months, draws us a slowly evolving seasonal
effect, a fairly smooth trend, and what is left over. This is pure
data-analysis, no explicit modelling assumptions are being made, we are
just applying a standard time-series algorithm called STL (LoESS) to let
the data speak for itself using the algorithm's default settings
for monthly time-series data. STL stands for ``Seasonal and Trend
decomposition using LoESS (locally estimated scatterplot smoothing)".
What do we see? Up to summer 2004 it is getting continuously busier and
busier. The number of patients being treated in Emergency almost doubles
from about 400 per month to about 800 per month. One may wonder if the
number of nurses in ED also doubled during this period -- it's highly
doubtful. The ``too small'' HGH was struggling to fight off closure
threats, staff was working harder and harder to keep it open. Then, the
number collapses. Possibly due to the situation which arose after Ben's
arrest and trial, potential patients tended to go elsewhere, if they had
the choice; but more likely, the policy of local health authorities was
dramatically changed, too. 25 miles away
is the very big teaching hospital Oxford Radcliffe, and in fact
Banbury's Horton General is part of the Radcliffe NHS trust group of
hospitals (nowadays called the Oxford University Hospitals NHS
Foundation Trust). For whatever reason, the number subsides to 500 --
600 per month. As mentioned before, there is a sudden dip in the very
last month, but that is meaningless: the last month was not quite over
when the data was submitted. The analysis should be redone without that
observation (but when we do that, nothing substantial changes).

The seasonal effect shows a strong annual spike downwards. It's the
Januaries! It is well known that in the Northern hemisphere, everywhere
where there really is a ``winter'', people simply stay home, and in
particular, don't go to hospital if they can help it, in January. There
are, for instance, less car accidents than in any other month,
because fewer people are out on the roads. Old people avoid
slippery paths by staying at home. Apart from that we don't see any
patterns. Accidents happen, and medical emergencies happen, at pretty
constant rate during the year. Correcting months for their varying
numbers of days (remember February) makes no discernible difference.

What is clear is that the number of people coming to that
hospital was steadily increasing in the years before Ben's troubles.
What about the numbers of nurses, of beds, of consultants? We plainly
see the amplitude of the monthly deviations from what you would expect
based on smooth annual trend and smooth seasonal average, increasing
with the overall scale. The bigger the overall expected number, the
bigger the random variation. This means in particular that at the time
which interests us most (winter 2003 -- 2004) the random variation is
largest!

Let's look (Figure \protect\hyperlink{figure10}{10}) at the total numbers
of transfers to Critical Care (i.e., the patient is no longer waiting in
the corridors for someone to make a decision, but is actually put in a
hospital bed in an intensive care ward) from Emergency. We have the
numbers with the interesting diagnoses of cardio-respiratory,
respiratory, and hypoglycaemic arrest, but we don't know if the
``arrest'' had been diagnosed before the patient was brought to the
hospital, or if it only occurred while the patient was waiting at
Emergency. Very sick patients who have to wait a long time in Emergency
before anyone can do anything with them are likely to suddenly get a lot
worse while they are waiting.

\noindent\protect\hypertarget{figure10}{}{}\includegraphics[width=14cm]{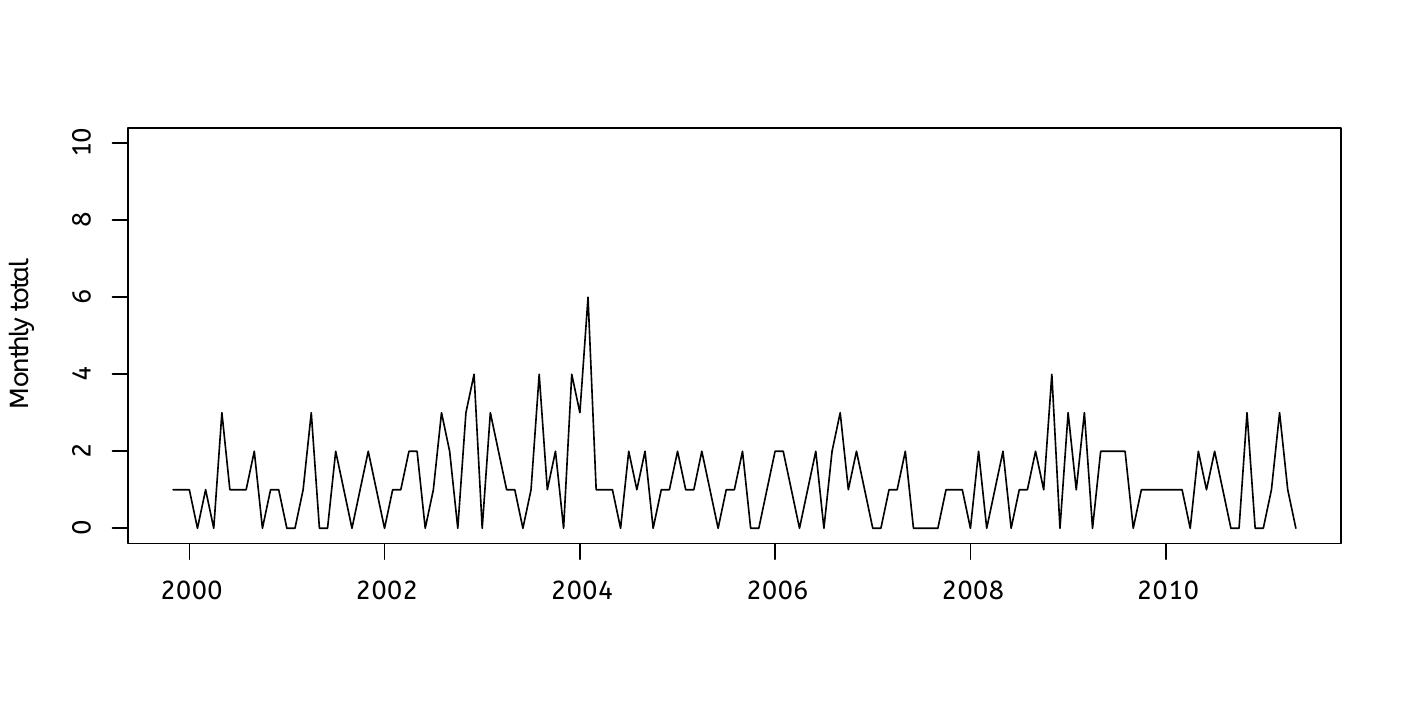}

\vspace*{-0.7cm}

\noindent\textbf{Figure 10.} Admissions from ED to CC with CR, Resp or Hypo arrest

\bigskip

What do we see? Just what we would expect, given the total numbers of
admissions which we just studied. A slow increase, then a collapse to a
stable, lower number. Take a look at what happens if we normalise the numbers by looking
at monthly totals per 100 admissions to ED, (Figure
\protect\hyperlink{figure11}{11}). The picture is much the same,
the differences over time are smaller. December 2003 and January 2004 are quite typical, but
February 2004 is noticeably high. But Ben is mostly absent in February 2004!
The next big spike is in November 2008. Both of those spikes coincide with
local peak levels of the monthly numbers
of admissions to Emergency.

\noindent\protect\hypertarget{figure11}{}{}\includegraphics[width=14cm]{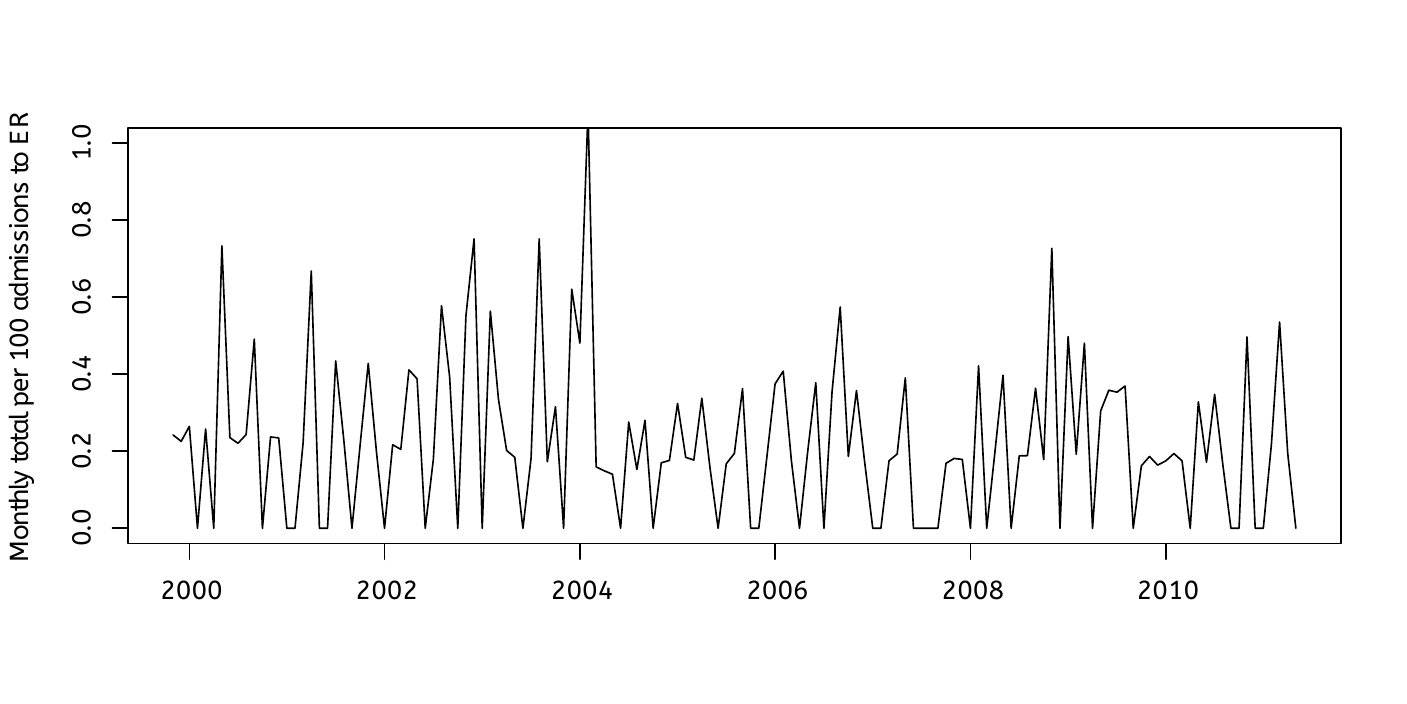}

\vspace*{-0.7cm}

\noindent\textbf{Figure 11.} Admissions from ED to CC with CR, Resp or Hypo arrest per 100
admissions to ED

\bigskip

In our opinion, the only thing that is unusual in
those months December 2003 -- February 2004 is that events are being
classified as \emph{respiratory} arrest instead of as
\emph{cardio-respiratory} or \emph{hypoglycaemic}. Why does the number
of cardio-respiratory arrests fall so suddenly in those three months? Do
we really believe that there were \emph{no} cardio-respiratory arrests
in December 2004? When were those numbers ``fixed'' in the official
records: day by day as patients were admitted? Or retrospectively after
Ben's case started, 5 February 2004?

The prosecution case would be that the respiratory arrests were
``extra'' events which occurred in ED through Ben's deliberate actions;
the patients in question perhaps came to ED because of an earlier
cardio-respiratory arrest. Against this is the fact that no-one ever saw
Ben doing anything at all which he shouldn't have been doing. One has to
go into the medical evidence. There are original hospital records which
are sketchy and difficult but not impossible for a layperson to
interpret. The medical experts for the prosecution are in no agreement
at all as to what caused what collapse, except that they are willing to
see something strange about each one, and willing to argue that Ben
\emph{could} have been to blame. The defence experts have sound reasons
to reject many of the hypothetical stories of the prosecution. Together, 
the expert opinions presented by both prosecution and defence witnesses do
allow an educated non-medical specialist to form their own opinion (after all,
the experts are writing in order to communicate to lawyers and to jurors). 

The extensive dossiers concerning the 18 selected patients were debated at
length during the trial. The jury decided that Ben had deliberately
harmed 17 patients, and in two cases, that this caused their subsequent deaths.

\section{Insights from forensic psychology and criminology}\label{criminology}

There is a major literature on the topic of ``Health Care Serial
Killers'' (HCSKs) in the fields of criminology and forensic psychology.
See especially \citep{yardley-wilson}. In those fields one studies
the characteristics of the kinds of persons who commit those crimes. More precisely,
of those who are convicted for those crimes.
Their material is necessarily mainly the evidence supplied by the winning
party: the judges' summing up or advice to jurors, media reports of the
prosecution's summary of the prosecution case. This means that scientists in these fields 
generally study past cases which led to convictions.  Different authors use
different definitions of \emph{serial} killer. A few authors explicitly
include ``suspicion of'' without ``conviction for''. Ben Geen was only convicted for two murders
but he allegedly harmed many more, and he is always included in the catalogues. 
Daniela Poggiali in Italy was convicted (and later, acquitted)
of only two murders, but suspected of a great many more.
Till her acquittal she also appeared on the lists.

If past cases are restricted to, say the last five decades in Western nations, the list of
such cases is pretty small -- perhaps 20 to 30. 
The list of ``certified'' HCSKs presented in various publications
included for a while the case of Lucia de Berk. Some of the pieces of
evidence which led to her conviction contributed to, and still stand in,
the current list of ``HCSK red flags'': warning signs, which managers in
health care should keep note of, and which are used as evidence against
new suspects in new cases when scientific evidence is given by
psychologists or profilers. For instance, ``thought by their colleagues
to be a weird person'' is one of the red flags. Another is,
``fascination with death''. Lucia had a couple of Stephen King novels at
home. ``Steals medications from the hospital''. Lucia had some empty
syringes at home which she had taken home for her daughter to use at
primary school in a ``show and tell'' presentation of what her Mum's day
job was. 

Most catalogues include Daniela Poggiali, Ben Geen, and Colin Norris.
The latter is another British nurse and it seems that his case will soon be
reopened due to new scientific insight into hypoglycaemic arrest (the CCRC has
already advised reopening).
He was arrested at around the same time as Ben Geen, similarly at the
peak of unrest concerning the then recent Shipman case.

Due to the Lucia de B.\ case, the ``red flag'' of
``increased incidence of deaths when the nurse is on duty'' has been
labelled as an unreliable indicator. 

It is interesting to look at the ``modus operandi'' of alleged HCSK's
who were convicted without powerful witness evidence of dangerous and illegal
behaviour. One of the ``weapons'' of choice appears to be Potassium
Chloride poisoning. The substance is readily available in hospitals,
though sometimes not sufficiently under lock and key. Put some into the
IV drip of a patient close to death and they will die very rapidly of
heart failure. Yet, a day or two later, a post-mortem toxicological
investigation will not be able to determine that that is what that
patient died of. If the drip has been used by many nurses and its
contents changed from time to time, it will be impossible to establish
that it was ever tampered with. It seems to us no coincidence that many dubious cases
revolve around K\(^{+}\)Cl\(^{-}\) poisoning. One might say it is a red
flag for a false accusation and for prosecutions which succeeded though
the accused was innocent.

Despite doubts as to the validity of these catalogues of serial killer nurses, 
these publications are extremely valuable. It is agreed that these cases are incredibly
rare. The incidence is, very roughly, 1 in 2 million per nurse per year, 
see \citep{forrest}, in which paper one finds a rough 1 in 1 million for nurses and doctors
together; both about equally frequent in the catalogue compiled in that paper. 
Compare this with the 1 or 2 thousand annual deaths in Dutch hospitals by \emph{admitted}
medical errors. We will now have a look at how such information might
impact reasoning through a simple Bayesian analysis of a hugely
simplified case.

\section{Bayesian analysis of a stylized case}\label{bayesian}

Convincing Bayesian analyses of these cases are conspicuously absent
from the academic literature. The landmark paper \citep{fienberg-kaye} 
announced that a Part 2 would be devoted to the Bayesian approach,
but Part 2 never materialized. A comprehensive 2016 review of the
use of Bayes in the law \citep{fenton-neil-berger} described the impediments against its success
but also the small number of cases where it has been used in court proceedings.

The Dutch econometrician Aart de Vos published a rather nice Bayesian
analysis of the case of Lucia de Berk in a Dutch language legal monthly journal.
An English translation prepared by Gill can be found at \url{https://gill1109.com/2021/05/24/condemned-by-statisticians/}.
It should be noted that his analysis uses some ``guessed'' numbers concerning the
analysis earlier mentioned by Elffers (which at the time was not public). In view of the now known errors in
the data supplied to the courts by the hospitals, his analysis would now come out even more favourable to the defence hypothesis.

What is clear, is that extraordinary claims require extraordinarily strong
evidence to support them. For a Bayesian, the prior probability that an
otherwise not particularly special nurse has been carrying out
deliberate murders of numerous patients in recent years, must be very
small indeed. Hence, in order to overcome that prior expectation, we
must have very strong evidence indeed. What about the evidence of a
stunningly small p-value in some study comparing death rate when a
particular nurse is present with death-rate when they are not present?

It is crucial that such statistical evidence has been generated in a professional way,
and this requires intensive collaboration between experienced statistical and medical experts.
First of all, discussions in a multi-disciplinary team are needed to
agree on definitions, criteria, a list of questions to be asked. Good statistical
methods must be used to obtain data and to assess its quality. Assumptions needed
for any statistical methodology, whether frequentist or Bayesian, must be
stated explicitly, and if possible evaluated. Any statisticians involved
must have deep understanding of the context. To give an example:
times of death reported on death certificates are in actual fact usually  
the times that a certified medical doctor ascertained that a patient
had already died. In a hospital, these times are often the times of the
hand-over between shifts. A nurse who clocks in early and clocks out
late can appear to be present at many more deaths than other nurses,
cf.\ the Poggiali case \citep{dotto-etal}.

Absence of hard evidence for actual malevolent
activities of the nurse in question is evidence of absence of such
activities. It is easy to illustrate this with a simple Bayes net model.
A nice tutorial introduction for legal scholars is \citep{ibs}.
Alternatively, in the appendix to this paper we analyse the model
``by hand''. It is so simple that all the calculations can be done with pen and paper, or on
a spreadsheet, or with the help of any basic computer language, from first principles.

Here is our graphical model (Bayes net) of a stylised and simplified serial killer nurse case.

\bigskip

\noindent\centerline{\includegraphics[width=3.2in]{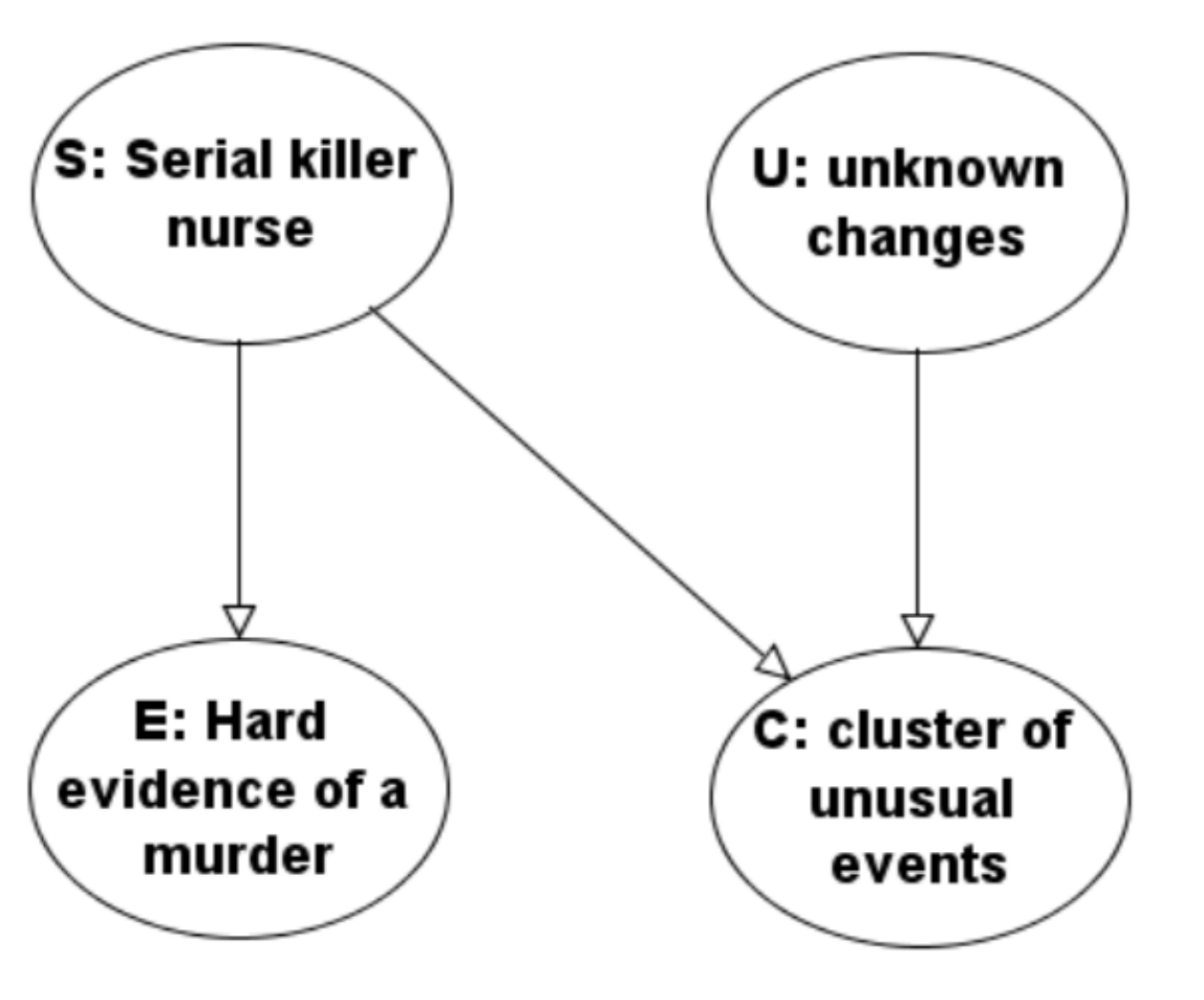}}
\noindent\textbf{Figure 4.} Simple illustrative Bayes net model

\bigskip

Our model is built around the Bayes net (DAG -- directed acyclic graph) of just four nodes shown in Figure 4.
The four nodes will represent binary random variables,  taking the values ``\texttt{yes}'' and ``\texttt{no}''.
The arrows between some of them represent a possible \emph{direct}
causal influence. It is always possible to order the nodes of a DAG, such that all arrows point forwards.
One of such orderings in our model is ($S, U, C, E$).
Having ordered the nodes so that all arrows point forward, the model says that the conditional probability
distribution of each variable given the values of all its predecessors (predecessors according to that ordering) only depends on the values of the predecessors
which are directly linked to it. Thus our model says that one could simulate realisations
of this $4$-tuple as follows: first of all, select an outcome for $S$ according to the marginal distribution of this random variable. 
Next, since there is no arrow from $S$ to $U$, draw an outcome for $U$ according to its marginal distribution. Next, since there are
arrows from both $S$ and $U$ to $C$, draw an outcome for $C$ according to its conditional distribution given the values of both 
these variables. Finally, draw an outcome for $E$ according to its conditional distribution given just the outcome of $S$.
The complete model therefore consists of the graph together with the sets of conditional probability distributions of
each variable conditional on all possible sets of values of the variables assumed to directly influence it, so-called
CPTs (Conditional Probability Tables). In the case of $S$ and $U$ these are actually tables of marginal probabilities.

The previous description is equivalent to another, which perhaps appeals more to classical ideas of causality.
Imagine deterministic mechanisms according to which the outcome of each variable is a completely deterministic function
of the variables which are supposed to directly influence it, together with a further random disturbance or latent variable.
All the latent variables are supposed to be statistically independent of one another,
and their probability distributions can be anything.
Simulate realisations of $(S, U, C, E)$ by first simulating an outcome of each random disturbance, four of them in our example.
Next, pick one of the nodes which has no arrows pointing to it. Its outcome is therefore a function just of
its own disturbance variable. After that, again and again, pick a node such that the outcomes of all nodes 
directly pointing at it have already been fixed. The outcome of the corresponding variable is a function of those outcomes together
with the outcome of its own (independent, local) disturbance.

Our graph has two root nodes, denoted by $S$ for serial killer and
$U$ for unknown changes. Examples of unknown changes could be changes 
in hospital policies and practice, such as changes in
protocols determining transfer of patients from one section to another,
introduction of new medical apparatus, changing milk formula for babies,
changing suppliers of medications. The environment can also change:
there may be new and not yet noticed viral, bacterial or fungal contamination 
somewhere in the hospital; the patient population may be changing for
all kinds of reasons quite outside of the hospital, such as changes in
the accessibility of alternative health care providers.
All kinds of things are of course changing all the time; the variable $U$
stands for changes which could cause an increase in the rate of
bad patient outcomes of some kind.
$S$ and $U$ are assume to be statistically independent.
Each can can trigger a surprising
cluster of cases associated with a particular nurse, either because they
(he or she) truly are a serial killer, or coincidentally because they
are often on duty at unfortunate times of day, or have longer shifts
than other nurses, and because they attract attention by odd behaviour.
Suppose that the prior probability of $S$ = ``yes'' is $10^{- 6}$ and
the prior probability of $U$ =``yes'' is $10^{- 2}$. 

The outcomes of these two variables are assumed to have direct probabilistic influence on the nodes linked
directly to them (known as their graph children). These are $C$ standing for a cluster of unusual and
unhappy events appearing to involve our nurse, and  $E$ standing for strong and direct evidence of the murder of one
particular patient by the nurse of interest. We allow that the probability of a cluster of
events given whether a serial killer is active and given whether there have been other unknown changes in the hospital,
may depend on the status of both graph parents $U$ and $S$. However, the conditional probability of finding (if one looks for it) direct evidence of
one (or more) actual murders by the nurse in question depends only on $S$, whether or not a serial killer nurse is active.
It doesn't depend on whether or not unknown changes to hospital procedures took place, $U$. 

This means that a total of six conditional probabilties 
of outcomes of graph children given the outcomes of their graph parents need to be written down to specify how the root nodes influence
the occurrence or not of a cluster of events and the finding (if it is
looked for) of direct evidence of one or more particular murders by our suspect. Four for $C$ and two for $E$.

The most important conditional probabilities concerning the cluster of events $C$
are the probability of $C$ = ``no''
conditional on $S$ = ``yes'' and $U$ = ``no'', which we take to equal
$10^{- 3}$,  the probability of $C$ = ``yes'' conditional on $S$ =
``no'' and $U$ = ``yes'', which we take to equal $10^{- 2}$, and the probability of $C$ = ``yes'' conditional on
$S$ = ``no'' and $U$ = ``no'' which we will take to be $10^{- 5}$. The first of these says that a serial killer almost
certainly triggers a cluster of cases. Unknown factors have some chance of triggering 
the observation of a cluster of cases, but the chance is small. If there are 
no unknown changes in the hospital and no serial killer the chance of a cluster is very small.
The remaining entry of theConditional Probability Table for node $C$ 
given its parents $S$ and $U$ will be taken as follows:  the probability of $C$ = ``no'' conditional on
$S$ = ``yes'' and $U$ = ``yes'' will be taken to be $10^{- 4}$. Since the conditioning eventuality is almost impossible,
(even relative to all the other unlikely events under consideration), this value is pretty irrelevant to later computations.

Regarding the node $E$, its only graph parent is $S$. We take the probability
that $E$ = ``no'' conditional on $S$
= ``yes'' to be $10^{- 2}$ while the probability that $E$ = ``yes''
conditional on $S$ = ``no'' will be $10^{- 3}$. Thus a serial killer is
rather likely to generate or leave \emph{some} hard evidence themselves
through evidence concerning one patient. On the other hand, a non serial killer nurse is unlikely to be
caught red-handed killing just one patient: the prior probability of murder by
health care professionals is quite low (though not so low as the probability
of serial murder).

The graph structure says that knowing also $U$ would not change
the probabilities of different values of $E$.

Either by hand, or by using standard software for computations with graphical models,
we can now perform numerous probability calculations.
Below we show the results obtained when the graphical model, including the conditional probability tables
we just specified, have been entered into the software program AgenaRisk.
Exactly the same results are obtained with other popular tools such as GeNIe
or Hugin, and in the statistical programming language R.

We can first of all compute the implied prior marginal probabilities of a cluster of
cases and of evidence implicating a particular nurse in murdering one particular patient, see Figure 5. 

\bigskip

\noindent\centerline{\includegraphics[width=3.2in]{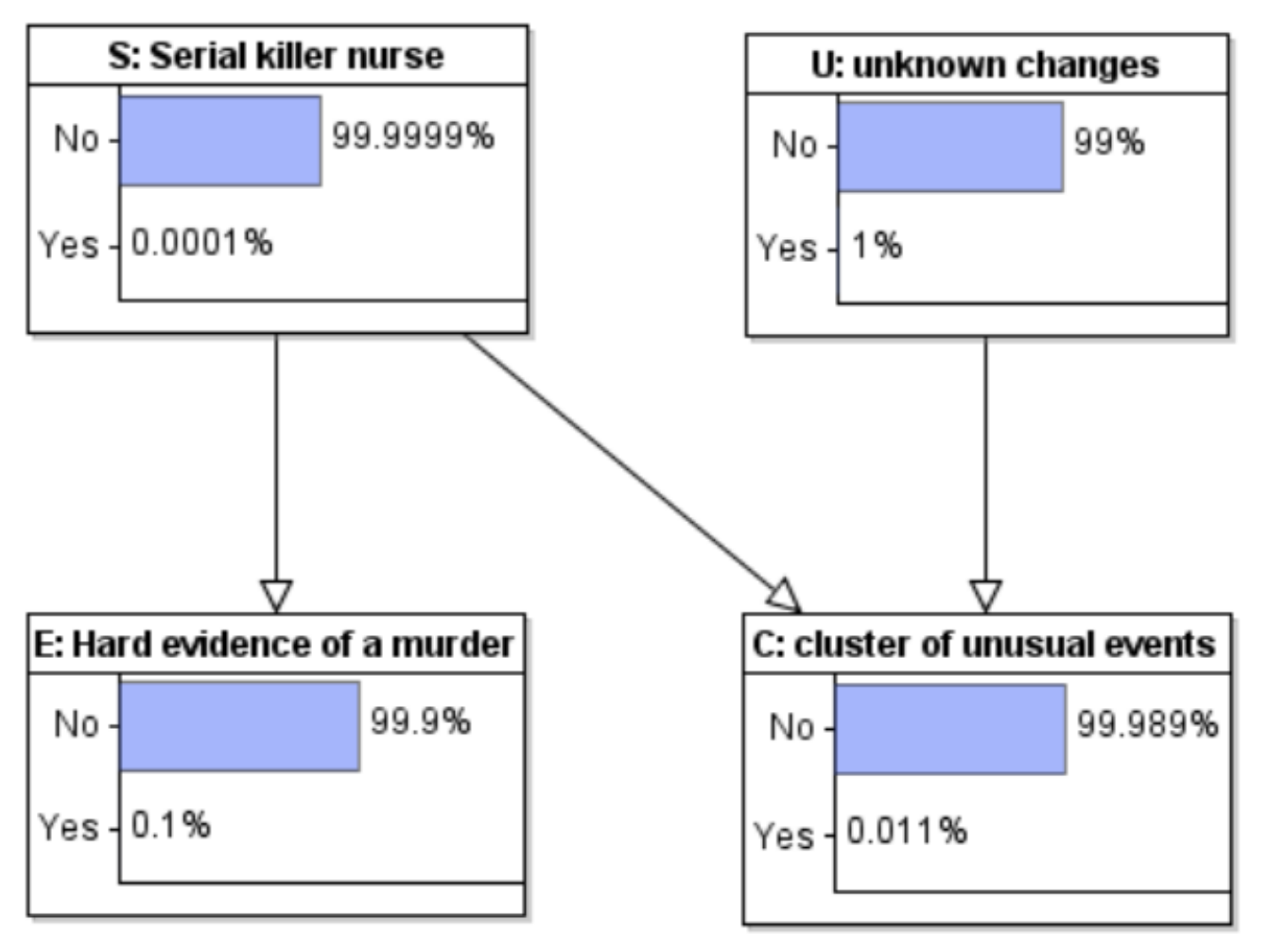}}
\noindent\textbf{Figure 5.} Model with marginal probabilities computed.

\bigskip

Both are rather small. There is a 0.011\% chance of an unexplained
cluster of events and a 0.1\% chance of incriminating evidence
concerning one patient and the nurse. 
Conditional on neither, the prior marginal probabilities for both $S$ and $U$ are
of course unchanged.

\bigskip

\noindent\centerline{\includegraphics[width=3.2in]{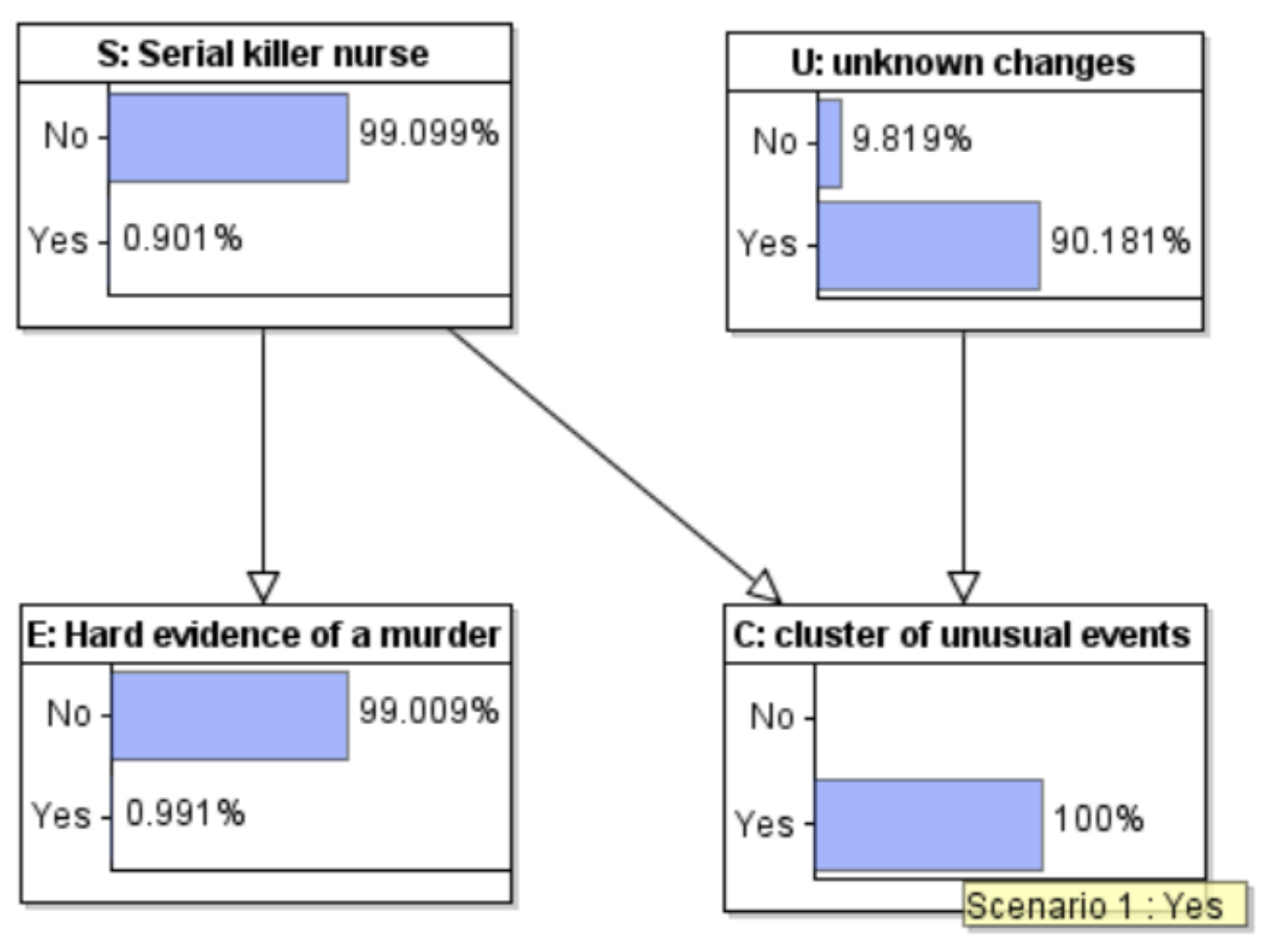}}
\noindent\textbf{Figure 6.}  Model with observations entered: Cluster of unusual events observed.

\bigskip

When we observe first (and so far, only) the cluster of unusual events (Figure 6) the
posterior probability of $U$ increases to over 90\%, while the probability
of $S$ increases to just 0.9\%. In other words, before we have looked for 
patient level evidence of murder, by far the most likely
explanation for the cluster of unusual events are unknown changes.

\bigskip

\noindent\centerline{\includegraphics[width=3.2in]{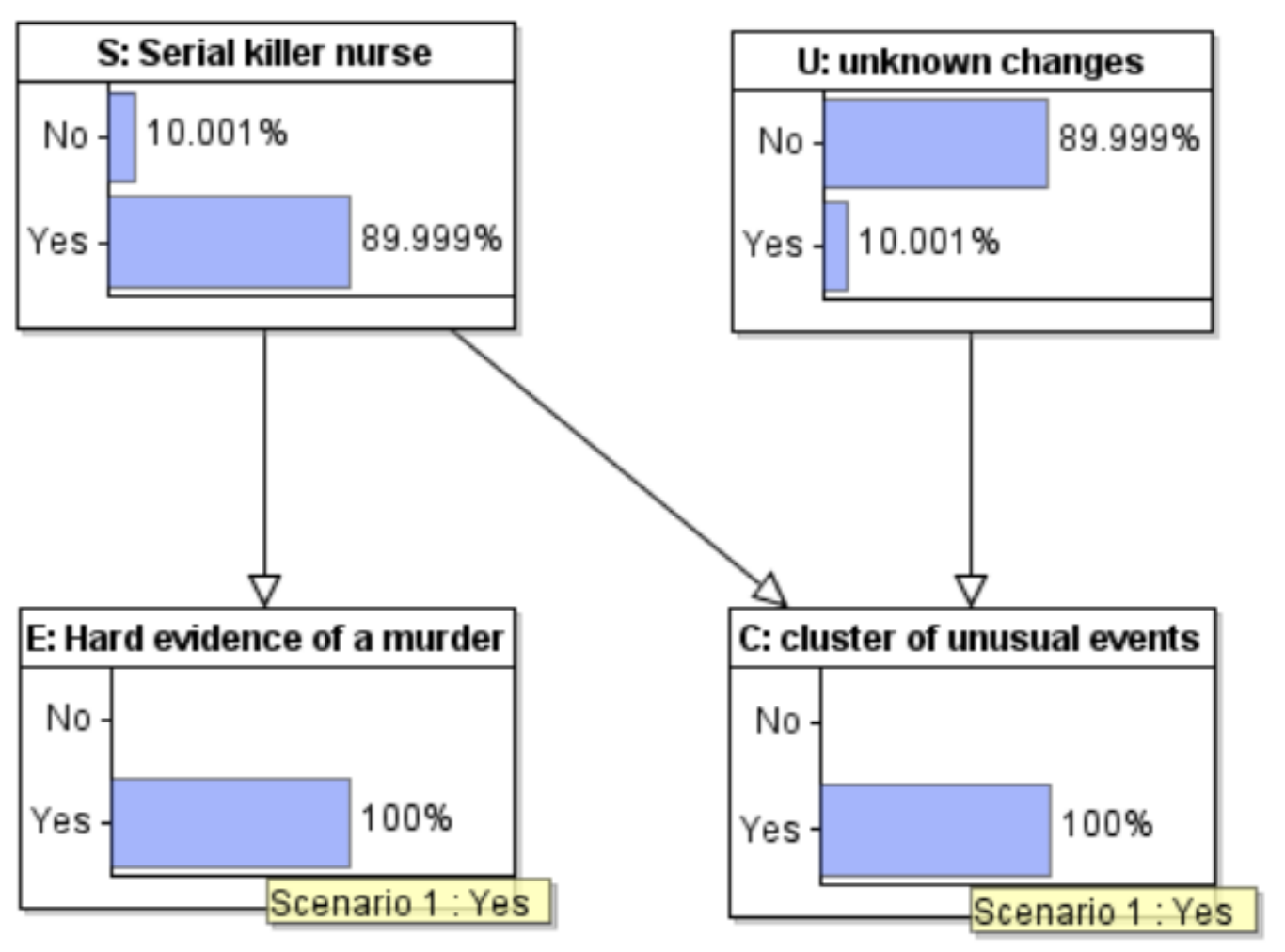}}
\noindent\textbf{Figure 7.}  Model with observations entered: Hard evidence also observed.

\bigskip

When both the cluster of events and incriminating single patient
evidence are observed  (Figure 7), the probability that the nurse is a serial killer
increases to just under 90\% (while the probability of unknown changes
drops to 10\%). \emph{So with these prior probabilities, even
given the strongest evidence possible for just one patient, there is still doubt that
that the nurse was a serial killer}.

In the Lucia de Berk case, the judges actually followed a sequential
form of this reasoning (though entirely verbal). There was not just one
patient for which hard evidence appeared to show that Lucia had caused
it harm. There was a sequence of three, in each case, the ``hard''
evidence was less hard, but each one tipped the scales further. After
the judges had concluded that Lucia was responsible for three deaths
they took it that she was also responsible for many more in the cluster
under investigation. The problem was not the argumentation, but the
evidence. The hard medical evidence was not hard at all. On careful
inspection and further toxicological measurements, it collapsed. 
Moreover, the ``medical'' argument that subsequent events were ``suspicious'' was
merely, in several cases, explicitly (in the words of a medical
expert witness) ``because Lucia was present at so many incidents''.

On the other hand, suppose we observe a cluster of events but then fail
to find strong evidence that the nurse hurt any one particular patient: now
the posterior probability that an unknown unknown caused the ``coincidence'' is
about 90\%; the probability that there was a serial killer is close to 0.

\bigskip

\noindent\centerline{\includegraphics[width=3.2in]{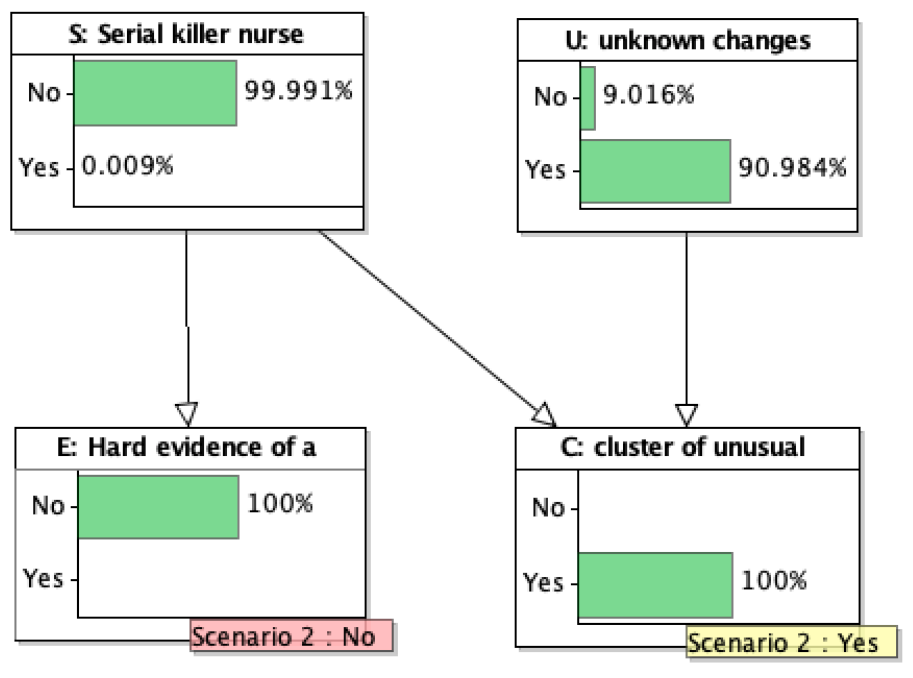}}
\noindent\textbf{Figure 8.}  Model with observations entered: Hard evidence observed absent.

\bigskip

This example is meant to be illustrative. It is not a serious model for
any specific case. In real cases, the interaction between observation of a surprising cluster of events
and investigation of specific cases occurs in many progressive stages, not just in one step. 
The value of the present model is that it could in principle be extended in order to model more features of
any actual case.
The example does show that absence of evidence can provide
very strong evidence of absence. All depends on the relative sizes of many very small
probabilities. We do not find our choices of those probabilities broadly unrealistic.
Their relative orders of magnitude determine the qualitative conclusions. 
If anything, some of the crucial numbers are unfavourable to our fictive nurse:
the base-rate for serial murder by nurses
is arguably twice as small as the number which we took. 
Past cases show that the incidence of ``unknown unknowns'' is not infinitesimal.
Recently in the UK, a striking cluster of events at one children's hospital was found
to be linked to a change in the supplier of baby formula milk. This was discovered by
an external statistician working for an official enquiry; hospital authorities had not thought of it before
and not suggested it as a possible cause.
We do not specify what we mean
by ``hard evidence for one murder''. From a subjective Bayesian point of view, 
what we mean by it is expressed in the relative sizes of the probabilities we put 
into the CPT's.

In the Lucia case, it was later discovered that a rise of deaths and other incidents in a medium care ward
was probably linked to a change in hospital policy concerning transfer from intensive care to medium care
of infants suffering from essentially incurable birth defects. The policy was supposed
to allow less babies to die in the hospital and more at home, however, it seems to have backfired. 
The policy was only known to a few top managers, and once they had committed
the whole hospital to accusations of serial murder by a particular nurse,
it was not going to be rapidly revealed. No defence lawyer could ever guess it, no investigative
journalist would uncover it.

Each specific case will have many more details and complicating factors. The
important point is that in a not negligible proportion of cases, there
is almost no hard evidence against the accused. The case has been
brought against them after a very long investigation, partly or entirely performed by
doctors and managers at the hospital where the events happened -- not by
independent forensic investigators -- and often hinges on some of the
``red flag'' items mentioned in the previous section. Gossip of
colleagues; some history of the nurse having once sought psychotherapy;
unusual personal character, perhaps connected to a traumatic childhood;
some past minor transgressions.

The reader who is unhappy with our choices of paramaters in this model is invited to perform
their own sensitivity analyses, and to build their own intuition, by playing with the parameters of
our model. We have built an online app: \url{https://nurse.staging.agenarisk.app/}.
It requires a password, please contact the corresponding author.
The basic network can naturally be run using any of the standard special software packages
for Bayes nets and causal modelling (GeNIe, Hugin, AgenaRisk), but it can also be run in
the general purpose statistical computer language R. The model is so simple that 
everything can also be done by a spreadsheet or by a simple computer program
written in any general purpose computer language. We provide code in R both for
direct calculations and using R's facilities for graphical models in the appendix, Section \ref{appendix}.

\section{Confirmational bias and other biases}\label{confirmation}

The investigation in the Ben Geen case appears to have suffered from \emph{confirmation
bias}, one of the most well-documented cognitive biases in psychology. Confirmation bias can also be thought of
as a blanket term for numerous specific kinds of statistical bias which can easily afflict
carelessly planned or executed empirical research of an observational nature. A criminal investigation is as an observational study,
in that one studies past events in order to identify the causes behind them.
Contrast this with an experimental study in which one intervenes in some phenomenon and observes what
are the consequences in the future of an intervention now.

We emphasize that we have absolutely no doubt that in the Ben Geen case, the hospital's own investigative team
and later police investigators were all acting with the best of possible intentions.
The same is true in the cases of Lucia de Berk and of Daniela Poggiali.
That is exactly the danger of confirmation bias: it is so easy to be totally unaware that an investigation is suffering from that bias.
It is so common, that ``standard procedures'' might easily reinforce it.

Because of its prevalence, researchers in medicine and epidemiology have studied and identified many varieties of bias which can easily
occur in observational studies. Moreover, they have identified research procedures which minimise or
in other ways account for possible bias. Nowadays, \emph{evidence based science} demands that evaluation of
new treatments for diseases, which also entails determining whether or not they have adverse effects on some
or many patients, is based on the gold standard of a randomized controlled clinical trial. (It has also become very
clear that very large samples are needed in order to gain useful evidence as to whether treatments are
beneficial or harmful, especially when this might depend on various patient characteristics: too small samples, and
misunderstanding of p-values, are parts of
the famous replication crisis afflicting especially the social sciences and the medical sciences).

In medicine there is a deep understanding of the problem of confounding factors, and particularly of hidden confounders: 
previously unanticipated factors which alter the chances of good or bad effects of a particular treatment.
A landmark paper in the field is \citep{sackett}: the author catalogues and names 35 different biases
which can occur in observational studies. Prof.~Jane Hutton's expert evidence in the Ben Geen case
(discounted as mere common sense by the judge) was based on the immense knowledge built among those involved in medical research
inspired by Sackett's paper. It was supported by the identical recommendations of a medical expert,
clinical physiologist Prof.\ David Denison; his evidence was also discounted by the judge.

The main messages from the medical field are the need for \emph{controls}
and the desirability of \emph{blinded} evaluation of whether any identified case was a ``treatment'' or 
a ``control'' case. Definitions (in this case, of the three kinds of adverse event; 
and on what counts as presence or absence of a nurse) need to be unambiguous,
defined in advance, and applied to all patients in the time period under study. What medical 
consequences befell each patient must be determined without knowledge of which nurses were 
involved in their care.

In a serial killer nurse case, where the nurse's presence is supposed to have caused bad outcomes for
patients in their care, the controls would be the cases in which similar bad outcomes occur though the nurse who is under suspicion
was not present. In the Ben Geen case it is clear that many collapses of patients occurred during Ben's absence,
but we do not even know exactly how many, nor how they were classified. The argument that there was an excess of normally rare respiratory arrests
when he was present is arguably based on hearsay. The lack of normally common cardio-respiratory events against a background of
overall normal total numbers is clear evidence that the alleged ``striking cluster'' was a striking cluster only in the imagination.

In psychology, confirmation bias is broadly defined as the subconsciously selective
gathering and weighting of evidence to support a specific hypothesis and the failure
to gather or recognize evidence that might count against that hypothesis
\citep{nickerson}. The bias has been shown to operate in many
investigative and forensic contexts \citep{dror-charlton, kassin-etal}. Summarizing a large body of empirical research,
\citep{lagnado} identifies three stages where confirmation bias can
derail investigations -- in the \emph{search}, \emph{evaluation} and
\emph{communication} of evidence -- with possible snowball effects when
these stages interact \citep{dror-etal, scherr-etal}. The investigations in both Geen and de Berk show confirmatory
biases at all three phases. For example, the investigative team in the
Ben Geen case only examined the patient records on Geen's shifts, but
not on shifts where he wasn't present. Without the comparison the
diagnostic value of the evidence against Geen cannot be fairly assessed
(as expert witnesses Jane Hutton and David Denison stated for the defence -- yet they were
disallowed by the judge to present this argument in court to the jury!).
Moreover, ambiguous or previously non-suspicious cases were
re-interpreted to fit with the hypothesis of Geen's guilt, and
alternative innocent explanations were neglected. Finally, the evidence
was communicated in a biased way, suppressing details of the
investigative process such that fact-finders at later stages, e.g., judge
and jury, were unaware that the evidence was selectively distorted.

These biases are particularly dangerous when a case involves complex
medical and statistical evidence, which many investigators and
fact-finders will not fully understand, and thus the potential for
misinterpretation is magnified. They are also exacerbated by people's
predilection to use stories to explain complex evidence \citep{pennington-hastie-a, pennington-hastie-b}. 
These stories typically focus on the actions of a
central protagonist, and a compelling story often overshadows a mass of
complex and equivocal evidence \citep{lagnado}.

So far, attempts to have the case of Ben Geen reconsidered by the CCRC have proved unsuccessful.
On the one hand, the story of the syringe grabs the public imagination. In our opinion,
a cunning health care serial killer does not walk into the hospital with a half-full
syringe of deadly poison in their pocket. Ben's story of why he emptied it, and his
girl-friend's corroboration of why he went back to hospital with it anyway, ring true
to us. This piece of evidence seems to us to be rather weak. The volume of
medical evidence is large but it seems to us ambiguous and inconclusive.
There is no hard medical evidence that any one of
those medical incidents was unnatural, despite intensive and biased search 
for such evidence. The jury were instructed to carry out an impossible
assignment, namely to decide each case on its own merits. Yet they
had heard the prosecution argument
that there was a highly unlikely cluster of unusual cases and that Ben was present
every time; this ``story'' was supported by some of the medical experts for the prosecution. 
Ben himself thought that the number of incidents on his shifts was disturbing.
But how could he have known what would have been a normal number?
He had worked at the hospital for a while in the past but this was his first winter there,
with, moreover, less supervision and more responsibilities, and much longer working hours.

By now, a number of authoritative statisticians have argued that the
conviction is unsafe due to the flaws in the hospital investigation. 
According to British lawyer Wendy Hesketh, 
who has studied several similar cases as well as this one \citep{hesketh}
\begin{quote}
\emph{The hospital's illegal and unqualified investigation team was only
looking for evidence to secure a conviction (Confirmation Bias) while
discarding or ignoring evidence that proved Ben's innocence. The
hospital's Serious Untoward Investigation Team initiated by Chief Nurse
Brock in her capacity of Executive Lead for Governance consisted of
several medical, nursing and medical records staff who were all
untrained in forensic investigative techniques, crime scene preservation
and the taking of witness statements. The team carried out an unlawful
and flawed investigation, the material from which was later presented to
medical experts appointed by the prosecution as legitimate. The opinion
of these medical experts was based on flawed evidence, which had been
given to those experts without their knowledge of how that evidence had
been obtained. Expert opinion given on the basis of ignorance of
improperly obtained evidence invalidates that medical expert evidence.
The judge and jury were not aware at the trial that the evidence had
been unlawfully obtained, nor of the risks to justice associated with it.}
\end{quote}

In fact, Hesketh wrote in the abstract of that paper ``A protocol for liaison between the police and the medical profession 
in dealing with crime committed by health professionals was implemented after being proposed 
in this journal [the \emph{Police Journal}] in 2003. However, rather than using their diverse skills to scrupulously investigate 
allegations of medico-crime, the police and health professions presume the accused is guilty and 
work together to prove this, disregarding any evidence to the contrary, thereby undermining the right to a fair trial. 
Whereas both were criticised in the past for failing to protect patients, 
they are likely to receive future criticism for contributing to serious miscarriages of justice.''

\section{Conclusions}\label{conclusions}
The case of Ben Geen bares remarkable similarities with the cases of Lucia de Berk and Daniela Poggiali \citep{dotto-etal}.
However, whereas the conviction of de Berk was overturned after review and reopening,
and that of Poggiali was overturned after cassation and reopening, 
all attempts to reopen the case against Geen have so far proven unsuccessful. 

A main purpose of the present
paper has been to document new statistical evidence in the Ben Geen case. There is more
that can be done: we need to obtain
staffing numbers, both of nurses and of medical specialists, at Horton General over the months and years around winter 2003--2004, in order to 
find out if Geen's claims that the pressure on the staff in the Accident and Emergency department was getting intolerable, 
have any support. We know that he was criticised for expressing such feelings to his superiors.
It would also be useful to find out what his working hours were during the three months 
in which he was accused of being on a killing spree.

Was the rate of those patient collapses during his
shifts really surprisingly large? We have some monthly totals and we know which particular patients were allegedly
harmed by him (and when), so we also know the number of ``arrests'' outside his shifts. The hospital investigation team allegedly investigated
around 30 incidents occurring during his shifts, but only brought about half of them to the attention of the police. One of those cases
failed to convince the jury. (The patient concerned had actually collapsed in the ambulance on the way to the hospital. This tells us that the hospital's
compilation of likely cases was not terribly accurate).

The total number of collapses of patients in A\&E waiting for further attention, during those
months, was also close to 30. This means that 10 or more collapses causing transfer to critical care occurred while Ben was absent. If he worked a great deal
of overtime, it could well be that the rate of these events was hardly different when he was on duty than when he was off duty, especially if he had a tendency
to be on duty at the busiest times of day. He was eager to obtain as much experience as possible. We certainly know that
the total number was actually typical for the time of year, and that it was the busiest time in the hospital  A\&E department that it had been for five or more years,
at the busiest time of year.

We think it is high time to systematically study the pattern of medical events in this case and build a Bayes net model incorporating statistical evidence, medical evidence,
and the evidence of the syringe. We hope some readers of this paper will be inspired to independently study the case.

For a systematic and comprehensive survey of lessons for the future which should be learnt from these
and many other cases, worldwide, we refer to the soon to be published hand-book
\citep{green-etal}. The opening words of that document are
worth reproducing here.

\begin{quote}\emph{Justice systems are sometimes called upon to evaluate cases in which health care professionals are suspected of killing their patients illegally.  These cases are difficult to evaluate because they involve at least two levels of uncertainty.  Commonly in a murder case it is clear that a homicide has occurred, and investigators must resolve uncertainty about who is responsible.  In the cases we examine here there is also uncertainty about whether homicide has occurred.   Investigators need to consider whether the deaths that prompted the investigation could plausibly have occurred for reasons other than homicide, in addition to considering whether, if homicide was indeed the cause, the person under suspicion is responsible.  
In this report, the RSS provides advice and guidance on the investigation and evaluation of such cases.  This report was prompted by concerns about the statistical challenges such cases pose for the legal system.  The cases often turn, in part, on statistical evidence that is difficult for lay people and even legal professionals to evaluate.  Furthermore, the statistical evidence may be distorted by biases, hidden or apparent, in the investigative process that render it misleading.  In this report the RSS provides advice and guidance on how to conduct investigations in such cases, with a particular focus on minimising the kinds of biases that could distort statistical evidence arising from the investigation.  This report also provides guidance on how to recognise and take account of such biases when evaluating statistical evidence and more broadly on how to understand the strengths and limitations of such evidence and give it proper weight.  
This report is designed specifically to help all professionals involved in investigating such cases and those who evaluate such cases in the legal system, including expert witnesses.  It will also be of interest to scholars and legal professionals who are interested in the role of statistics in evidentiary proof, and more generally to anyone interested in improving criminal investigations.}
\end{quote}

\vspace{6pt} 

\authorcontributions{The first author initiated this work; the second two authors contributed much further material on Bayes nets and on forensic psychology. All authors have read and agreed to the published version of the manuscript.}

\funding{This research received no external funding.}

\dataavailability{Data and R scripts are available at  \url{https://github.com/gill1109/bengeen}.}

\acknowledgments{Richard Gill thanks Mark van der Werf and Flavio Azevedo for computing assistance.}

\conflictsofinterest{Richard D. Gill worked largely pro bono as expert for Ben Geen's defence team during an initial attempt, 2014--2015, to have the case reviewed by the UK's CCRC. 
He was personally involved for many more years and with numerous other volunteers in attempts to
get the cases of Lucia de Berk, Daniela Poggiali, and Ben Geen reviewed and re-opened, leading to final non-guilty verdicts in two of the three cases.
Norman Fenton is an owner of the company AgenaRisk.}

\reftitle{References}


\section{Appendix}\label{appendix}

Explicit calculations for the simple Bayes net. Recall that our graphical model had four binary nodes. Two are possible \emph{causes}: $U$, ``unknown'', $S$, ``serial killer''.They are causes
of the effects $C$, ``cluster'', and $E$, ``evidence''.  In graphical model language, the four nodes of the graph stand for four binary random variables, all taking the values ``YES'' and ``NO''. For instance there either is, or is not, a cluster of events; our nurse either is, or is not, a serial killer. Some unknown factor either does or does not potentially cause a cluster of events, and hard evidence of one particular murder may or may not be found. 

In our model, $U$ and $S$ are statistically independent binary random variables. There are therefore four possible initial causal conditions. In particular, it is just possible that unknown factors are likely to cause a cluster of events \emph{and} that our nurse is a serial killer. To simply calculations ``by hand'', we will discard this extremely rare possibility. Changing notation, $O$, $U$ and $S$ will now stand for three mututally exclusive and exhaustive \emph{events}. The initial possibilities will be $O$ (``neither $U$ nor $S$''), $U$ and $S$. The new event "S" is the old event $\{S = \textrm{YES}\}$. The new event $O$ stands for \emph{nothing unusual is going on: no serial killer, no unknown change to the system likely to cause a cluster of events}.

$C$ and $E$ now stand for the events ``observation a cluster of events'', and ``observation of hard evidence for a murder''. We will write $\overline C$ and $\overline E$ for the events ``no cluster of events observed'', and ``hard evidence for a murder not found, despite search''.

Our model states that $C$ and $E$ are conditionally independent given $O$, $C$ and $U$, and that the probability of $E$ given $O$, $C$ or $U$ 
only depends on whether or not $S$ holds. The numbers we previously have for various probabilities and conditional probabilities in the Bayes net 
can now be summarized as
$$P(S) = s = 10^{-6}$$
$$P(U) = u = 10^{-2}$$
$$P(O) = o = 1 - u - s$$

$$P(\overline C | S) = c = 10^{-3}$$
$$P(C | U) = b = 10^{-2}$$
$$P(C | O) = a = 10^{-5}$$

$$P(\overline E | S) = d = 10^{-2}$$
$$P(E | \overline S) = e = 10^{-3}$$
Notice that by assuming the graph structure we have made a number of conditional independence assumptions, for instance:
$$P(C | S) = P(C | S, E) = P(C | S, \overline E);$$
given $S$, $C$ and $E$ are statistically independent. (This statement holds both with respect to our original random variable notation, and with respect to the new events notations).

Using the definition of conditional probability $P(A|B) = P(A\textrm{~and~}B)/P(B)$, one can
write down formulas for the conditional probabilities of the three possible causes given only observation of a cluster of events; then of observation of a cluster of events and of 
hard evidence for a murder; and last of all, of observation of a cluster of events and observation of no hard evidence for a murder despite searching for it.
The first three conditional probabilities represent the situation after the cluster has been observed but before evidence is looked for that foul play was involved. The next two groups of three
represent the two alternative situations after such a search has been carried out, depending on whether or not it turned up hard evidence.
$$P(O|C) = (1 - u - s)*a / ((1 - u - s)*a +u*b + s*(1 - c))$$
$$P(U|C) = u*b / ((1 - u - s)*a +u*b + s*(1 - c))$$
$$P(S|C) = s*c / ((1 - u - s)*a +u*b + s*(1 - c))$$

$$P(O|C,E) = (1 - u - s)*a*e / ((1 - u - s)*a*e + u*b*e +s*(1 - c)*(1 - d))$$
$$P(U|C,E) = u*b*e / ((1 - u - s)*a*(1 - e) + u*b*e + s*(1 - c)*(1 - d))$$
$$P(S|C,E) = s*c*(1 - d) / ((1 - u - s)*a*(1 - e) + u*b*e + s*(1 - c)*(1 - d))$$

$$P(O|C, \overline E) = (1 - u - s)*a*(1 - e) / ((1 - u - s)*a*(1 - e) + u*b*(1 - e) + s*(1 - c)*d)$$
$$P(U|C, \overline E) = u*b*(1 - e) / ((1 - u - s)*a*(1 - e) + u*b*(1 - e) + s*(1 - c)*d)$$
$$P(S|C, \overline E) = s*c*(1 - d) / ((1 - u - s)*a*(1 - e) + u*b*(1 - e) + s*(1 - c)*d)$$

\bigskip
\noindent \textbf{Here is code in the language \texttt{R} which calculates the probabilities of interest:}

\begin{verbatim}
s = 10^-6       # p(S)
u = 10^-2       # p(U)
o = 1 - u - s    #p(O)

cCOMP = 10^-3       # p(not C | S);    "COMP" stands for "complement
b = 10^-2       # p(C | U)
a = 10^-5       # p(C | O)

d = 10^-2       # p(not E | S)
e = 10^-3       # p(E | not S)

c = 1 - cCOMP        
dCOMP = 1 - d
eCOMP = 1 - e

p.O.C = o*a / (o*a +u*b + s*c)
p.U.C = u*b / (o*a +u*b + s*c)
p.S.C = s*c / (o*a +u*b + s*c)

p.O.CE = o*a*e / (o*a*e + u*b*e +s*c*dCOMP)
p.U.CE = u*b*e / (o*a*e + u*b*e + s*c*dCOMP)
p.S.CE = s*c*dCOMP / (o*a*e + u*b*e + s*c*dCOMP)

p.O.CnotE = o*a*eCOMP / (o*a*eCOMP + u*b*eCOMP + s*c*d)
p.U.CnotE = u*b*eCOMP / (o*a*eCOMP + u*b*eCOMP + s*c*d)
p.S.CnotE = s*c*d / (o*a*eCOMP + u*b*eCOMP + s*c*d)

print("Probs of O, U, S given C (%)")
round(100*c(p.O.C, p.U.C, p.S.C), 4)
sum(100*c(p.O.C, p.U.C, p.S.C))

print("Probs of O, U, S given C and E (%)")
round(100*c(p.O.CE, p.U.CE, p.S.CE), 4)
sum(100*c(p.O.CE, p.U.CE, p.S.CE))

print("Probs of O, U, S given C and not E (%)")
round(100*c(p.O.CnotE, p.U.CnotE, p.S.CnotE), 4)
sum(100*c(p.O.CE, p.U.CE, p.S.CE))
\end{verbatim}

\bigskip
\noindent \textbf{Here are the results of running that code:}

\begin{verbatim}
> print("Probs of O, U, S given C (%)")
> print("Probs of O, U, S given C (%)")
[1] "Probs of O, U, S given C (%)"
> round(100*c(p.O.C, p.U.C, p.S.C), 4)
[1]  8.9270 90.1721  0.9008
> sum(100*c(p.O.C, p.U.C, p.S.C))
[1] 100
> 
> print("Probs of O, U, S given C and E (%)")
[1] "Probs of O, U, S given C and E (%)"
> round(100*c(p.O.CE, p.U.CE, p.S.CE), 4)
[1]  0.9009  9.0999 89.9992
> sum(100*c(p.O.CE, p.U.CE, p.S.CE))
[1] 100
> 
> print("Probs of O, U, S given C and not E (%)")
[1] "Probs of O, U, S given C and not E (%)"
> round(100*c(p.O.CnotE, p.U.CnotE, p.S.CnotE), 4)
[1]  9.0074 90.9835  0.0091
> sum(100*c(p.O.CE, p.U.CE, p.S.CE))
[1] 100
\end{verbatim}

\bigskip

\noindent \textbf{The original graphical model can also be implemented entirely in R:}

\begin{verbatim}
library(Rgraphviz)
library(gRain)
yn <- c("yes", "no")
S <- cptable(~SerialKiller, values=c(1, 999999), levels = yn)
U <- cptable(~UnknownCause, values=c(1, 99), levels = yn)
E.S <- cptable(~EvidenceOneMurder:SerialKiller, values = c(99, 1, 1, 999), 
     levels = yn)
C.SU <- cptable(~ClusterofEvents:SerialKiller:UnknownCause, 
     values = c(999, 1, 1, 99, 999, 1, 1, 99999), levels = yn)
killerNurse <- compileCPT(S, U, C.SU, E.S)
killerNurse
killerNurse$SerialKiller
killerNurse$UnknownCause
killerNurse$EvidenceOneMurder
killerNurse$ClusterofEvents
killerNurse_bn <- grain(killerNurse)
killerNurse_bn <- propagate(killerNurse_bn)
killerNurse_bn
saveHuginNet(killerNurse_bn, "killerNurse.net")

#########################################################################
# No instantiation of evidence
querygrain(killerNurse_bn, nodes = c("SerialKiller", "UnknownCause"), 
     type = "joint")
round(100*querygrain(killerNurse_bn, 
     nodes = c("SerialKiller", "UnknownCause"), type = "joint"))

#########################################################################
# We observe no mysterious cluster and no evidence of a murder
killerNurse_bn0 <-  setEvidence(killerNurse_bn, 
     evidence = list(ClusterofEvents="no", EvidenceOneMurder="no"))
querygrain(killerNurse_bn0, nodes = c("SerialKiller", "UnknownCause"), 
     type = "joint")
round(100*querygrain(killerNurse_bn0, 
     nodes = c("SerialKiller", "UnknownCause"), type = "joint"))
# 99% chance nothing going on at all. 1% chance that things have changed 
# but we did not see evidence of it

#########################################################################
# We just observe a suspicious cluster of events involving our nurse
killerNurse_bn1 <-  setEvidence(killerNurse_bn, 
     evidence = list(ClusterofEvents="yes"))
querygrain(killerNurse_bn1, nodes = c("SerialKiller", "UnknownCause"), 
     type = "joint")
round(100*querygrain(killerNurse_bn1, 
     nodes = c("SerialKiller", "UnknownCause"), type = "joint"))
# 90% chance something changed but our nurse is not a serial killer.
# 9% chance that actually nothing changed at all
# 1% chance we have a serial killer on our hands

#########################################################################
# We observe a suspicious cluster and strong evidence of wrong doing 
# by our nurse on one particular patient
killerNurse_bn11 <-  setEvidence(killerNurse_bn, 
     evidence = list(ClusterofEvents="yes", EvidenceOneMurder="yes"))
querygrain(killerNurse_bn11, nodes = c("SerialKiller", "UnknownCause"), 
     type = "joint")
round(100*querygrain(killerNurse_bn11, 
     nodes = c("SerialKiller", "UnknownCause"),  type = "joint"))
# 90% chance our nurse is a serial killer
# 9% chance she is not a serial killer but 
# something else caused the cluster of events
# 1% chance nothing is going on at all; 1% chance 
# she's a serial killer and things changed

#########################################################################
# We observe a suspicious cluster but no strong evidence of wrong doing 
# by our nurse on any one patient
killerNurse_bn12 <-  setEvidence(killerNurse_bn,
     evidence = list(ClusterofEvents="yes", EvidenceOneMurder="no"))
querygrain(killerNurse_bn12, nodes = c("SerialKiller", "UnknownCause"), 
     type = "joint")
round(100*querygrain(killerNurse_bn12, 
     nodes = c("SerialKiller", "UnknownCause"), type = "joint"))
# 91% chance something changed, the nurse is not a serial killer
# 9% chance nothing going on at all
\end{verbatim}


\begin{thebibliography}{}

\bibitem[Dotto et al, 2022]{dotto-etal}
Dotto, F., Gill, R.D., and Mortera, J. (2022) Statistical Analyses in the
case of an Italian nurse accused of murdering patients. \emph{Law,
Probability and Risk} (to appear). \url{https://arxiv.org/abs/2202.08895}

\bibitem[Dror \& Charlton, 2006]{dror-charlton}
Dror, I. E., and Charlton, D. (2006). Why experts make errors. 
\emph{Journal of Forensic Identification} \textbf{56}(4), 600. \url{https://www.researchgate.net/publication/248440075_Why_Experts_Make_Errors}

\bibitem[Dror et al, 2017]{dror-etal}
Dror, I. E., Morgan, R. M., Rando, C., and Nakhaeizadeh, S. (2017). 
Letter to the editor—The bias snowball and the bias cascade effects:
Two distinct biases that may impact forensic decision making. 
\emph{Journal of forensic sciences} \textbf{62} (3), 832--833. 
\url{https://doi.org/10.1111/1556-4029.13496}

\bibitem[Elffers, 2002a]{elffersa}
Elffers, H. (2002a) \emph{Distribution of incidents of resuscitation and
death in the Juliana Kinderziekenhuis and the Rode Kruisziekenhuis}
(Unofficial English translation of Dutch original).
\url{https://www.math.leidenuniv.nl/~gill/Elffers1eng.pdf}

\bibitem[Elffers, 2002b]{elffersb}
Elffers, H. (2002b) \emph{Elaborated analysis. Distribution of incidents
of resuscitation and death at the Juliana Kinderziekenhuis and the Rode
Kruisziekenhuis} (Unofficial English translation of Dutch original).
\url{https://www.math.leidenuniv.nl/~gill/Elffers2eng.pdf}

\bibitem[Fenton et al, 2016]{fenton-neil-berger}
Fenton, N., Neil, M., and Berger, D. (2016).
Bayes and the Law,
\emph{Annual {R}eview of Statistics and Its {A}pplication},
\textbf {3}(1), 51--77.
\url{https://doi.org/10.1146/annurev-statistics-041715-033428}, 
\url{http://www.eecs.qmul.ac.uk/~norman/papers/bayes_and_the_law_revised_FINAL.pdf}

\bibitem[Fienberg \& Kaye, 1991]{fienberg-kaye}
Fienberg, S. E., and Kaye, D. H. (1991) Legal and Statistical Aspects of
Some Mysterious Clusters. \emph{Journal of the Royal Statistical Society
A} \textbf{154} (1), pp. 61--74 \url{https://ssrn.com/abstract=2777146}

\bibitem[Forrest, 1995]{forrest}
Forrest, A.R.W. (1995) Nurses who systematically harm their patients.
\emph{Medical Law International} \textbf{1}, pp. 411--421.
\url{https://doi.org/10.1177/096853329500100404}

\bibitem[Gill, 2014]{gill}
Gill, R.D. (2014) \emph{Rarity of respiratory arrest in ED?}
\url{https://arxiv.org/abs/1407.2731}

\bibitem[Gill et al, 2018]{gill-etal}
Gill, R.D., Groeneboom, P., and de Jong, P. (2018) Elementary Statistics
on Trial (the case of Lucia de Berk). \emph{Chance} \textbf{31} (4), pp.
9--15. \url{https://doi.org/10.1080/09332480.2018.1549809}

\bibitem[Green et al, 2022]{green-etal}
Green,~P., Gill, R.D., Mackenzie,~N., Mortera,~J., and Thompson,~W. (2022).
\emph{Healthcare Serial Killer or Coincidence?
Statistical Issues in Investigation of Suspected Medical Misconduct}.
Royal Statistical Society, Statistics and the Law section. London.

\bibitem[Hesketh, 2012]{hesketh}
Hesketh,~W. (2012) The police-health profession's protocol: a review.
\emph{The Police Journal} \textbf{85}, 203--220.
\url{https://doi.org/10.1350/pojo.2012.85.3.576}

\bibitem[Ibs, 2016]{ibs}
Ibs, I.C. (2016) \emph{Applications of Bayesian Networks In Legal Reasoning}.
Bachelor thesis, University of Osnabrück.
\url{https://www.researchgate.net/publication/311667346_Applications_of_Bayesian_Networks_In_Legal_Reasoning}

\bibitem[Kassen et al, 2013]{kassin-etal}
Kassin, S., Dror, I., and Kukucka, J. (2013). The Forensic Confirmation Bias: 
Problems, Perspectives, and Proposed Solutions. 
\emph{Journal of Applied Research in Memory and Cognition} \textbf{2}, 42--52.
\url{https://doi.org/10.1016/j.jarmac.2013.01.001}

\bibitem[Lagnado, 2021]{lagnado}
Lagnado, D. A. (2021). \emph{Explaining the Evidence: How the Mind Investigates the World}. 
Cambridge University Press.

\bibitem[Nickerson, 1998]{nickerson}
Nickerson, R. S. (1998). Confirmation bias: A ubiquitous phenomenon in many guises. 
\emph{Review of General Psychology} \textbf{2}(2), 175--220.
\url{https://doi.org/10.1037/1089-2680.2.2.175}

\bibitem[Pennington \& Hastie, 1986]{pennington-hastie-a}
Pennington, N., \& Hastie, R. (1986). Evidence evaluation in complex decision making. 
\emph{Journal of Personality and Social Psychology} \textbf{ 51}(2) 242--258. 
\url{https://doi.org/10.1037/0022-3514.51.2.242}

\bibitem[Pennington \& Hastie, 1992]{pennington-hastie-b}
Pennington, N., \& Hastie, R. (1992). Explaining the evidence: Tests of the Story Model for juror decision making. 
\emph{Journal of Personality and Social Psychology}, \textbf{62}(2) 189--206. \url{https://doi.org/10.1037/0022-3514.62.2.189}

\bibitem[Sackett, 1979]{sackett}
Sackett, D.L. (1979). Bias in analytical research. \emph{J. Chron. Dis.} \textbf{32}, 51--63.
\url{https://doi.org/10.1016/0021-9681(79)90012-2}

\bibitem[Scherr et al, 2013]{scherr-etal}
Scherr, K. C., Redlich, A. D., and Kassin, S. M. (2013). Cumulative disadvantage: 
A psychological framework for understanding how innocence can lead to confession, wrongful conviction, and beyond. 
\emph{Perspectives on Psychological Science} \textbf{15}(2), 353--383.
\url{https://doi.org/10.1177/1745691619896608}

\bibitem[Yardley \& Wilson, 2014]{yardley-wilson}
Yardley, E., and Wilson, D. (2014) In Search of the `Angels of Death':
Conceptualising the Contemporary Nurse Healthcare Serial Killer.
\emph{J. Investig. Psych. Offender Profil.}
\href{\%20https://doi.org/10.1002/jip.1434}{https://doi.org/10.1002/jip.1434}

\end{thebibliography}
\end{document}